\documentclass[aoas,MSNbibl,nameyear,seceqn,dvips]{arximspdf}
\usepackage{graphicx}

\doi{10.1214/10-AOAS383}
\volume{5}
\issue{1}
\pubyear{2011}
\firstpage{523}
\lastpage{550}

\makeatletter

\newtheorem{theorem}{Theorem}[section]
\newtheorem{prop}[theorem]{Proposition}
\newtheorem{cor}[theorem]{Corollary}
\newproclaim{example}{Example}
\def\eqref#1{(\ref{#1})}

\makeatother

\begin{document}
\begin{frontmatter}

\title{Spatial models generated by nested stochastic
partial differential equations, with an application to global ozone mapping}
\runtitle{Spatial models generated by nested SPDEs}

\begin{aug}
\author{\fnms{David} \snm{Bolin}\ead[label=e1]{bolin@maths.lth.se}}
\and
\author{\fnms{Finn} \snm{Lindgren}\ead[label=e2]{finn@maths.lth.se}}

\runauthor{D. Bolin and F. Lindgren}

\affiliation{Lund University}

\address{Mathematical Statistics\\
Centre for Mathematical Sciences\\
Lund University, Box 118\\
SE-22100 Lund\\ Sweden \\
\printead{e1}\\
\phantom{E-mail:\ }\printead*{e2}}
\end{aug}

\received{\smonth{2} \syear{2010}}
\revised{\smonth{7} \syear{2010}}

%
\begin{abstract}
A new class of stochastic field models is constructed using nested
stochastic partial differential equations (SPDEs). The model class is
computationally efficient, applicable to data on general smooth
manifolds, and includes both the Gaussian Mat\'{e}rn fields and a wide
family of fields with oscillating covariance functions. Nonstationary
covariance models are obtained by spatially varying the parameters in
the SPDEs, and the model parameters are estimated using direct
numerical optimization, which is more efficient than standard Markov
Chain Monte Carlo procedures. The model class is used to estimate daily
ozone maps using a large data set of spatially irregular global total
column ozone data.
\end{abstract}

%
\begin{keyword}
\kwd{Nested SPDEs}
\kwd{Mat\'{e}rn covariances}
\kwd{nonstationary covariances}
\kwd{total column ozone data}.
\end{keyword}

\end{frontmatter}

\section{Introduction}

Building models for spatial environmental data is a challenging
problem that has received much attention over the past years.
Nonstationary covariance models are often needed since the
traditional stationary assumption is too restrictive for
capturing the covariance structure in many problems.
Also, many environmental data sets today contain massive amounts
of measurements, which makes computational efficiency another
increasingly important model property. One such data set, which will be
analyzed in this work,
is the the Total Ozone Mapping Spectrometer (TOMS) atmospheric ozone
data [\citet{toms1}].
The data was collected by a TOMS instrument onboard the the near-polar,
Sun-synchronous orbiting satellite
Nimbus-7, launched by NASA on October 24, 1978. During the sunlit
portions of the satellite's orbit, the
instrument collected data in scans perpendicular to the orbital plane.
A new scan was started every eight seconds as the spacecraft moved from
south to north. A number of recent papers in the statistical literature
[\citet{cressie08}, \citet{jun08}, \citet{stein07}]
have studied the data, and it requires
nonstationary covariance structures as well as efficient computational
techniques due to the large number of observations.

A covariance model that is popular in environmental statistics
is the Mat\'{e}rn family of covariance functions [\citet{matern60}].
The Mat\'{e}rn covariance function has a shape parameter, $\nu$, a
scale parameter, $\kappa$, and a variance\footnote{With this
parametrization, the variance $C(\mathbf{0})$ is $\phi^2(4\pi)^{-
{d/2}}\Gamma(\nu)\Gamma(\nu+
{d/2})^{-1}\kappa^{-2\nu}$.} parameter, $\phi^2$, and can be
parametrized as
%
\begin{eqnarray}\label{paperC:eq:matern}
C(\mathbf{h}) = \frac{2^{1-\nu}\phi^2}{(4\pi)^{{d/2}}\Gamma(\nu+
{d/2})\kappa^{2\nu}}(\kappa\|\mathbf{h}\|)^{\nu}K_{\nu}(\kappa\|
\mathbf{h}\|),\qquad \mathbf{h}\in\mathbb{R}^d,
\end{eqnarray}
where $K_{\nu}$ is a modified Bessel function of the second kind of
order $\nu>0$. One drawback with defining the model directly through a
covariance function, such as \eqref{paperC:eq:matern}, is that it makes
nonstationary extensions difficult. Another drawback is that, unless
the covariance function has compact support, the computational
complexity for calculating the Kriging predictor based on $n$
measurements is $\mathcal{O}(n^3)$. This makes the Mat\'{e}rn covariance
model computationally infeasible for many environmental data sets.\par

Recently, \citet{lindgren10} derived a method for explicit, and
computationally efficient, Markov representations of the Mat\'{e}rn
covariance family. The method uses the fact that a random process on
$\mathbb{R}^d$ with a Mat\'{e}rn covariance function is a solution to the
stochastic partial differential equation (SPDE)
%
\begin{equation}\label{paperC:sde}
(\kappa^2-\Delta)^{{\alpha/2}}X(\mathbf{s}) = \phi\mathcal{W}
(\mathbf{s}),
\end{equation}
where $\mathcal{W}(\mathbf{s})$ is Gaussian white noise, $\Delta$ is
the Laplace
operator, and $\alpha= \nu+ d/2$ [\citet{whittle63}]. Instead of
defining Mat\'{e}rn fields through the covariance functions
\eqref{paperC:eq:matern}, \citet{lindgren10} used the solution
to the
SPDE \eqref{paperC:sde} as a definition. This definition is valid not
only on $\mathbb{R}^d$ but also on general smooth manifolds, such as
the sphere,
and facilitates nonstationary extensions by allowing the SPDE
parameters $\kappa^2$ and $\phi$ to vary with space. The Markov
representations
were obtained by considering approximate stochastic weak solutions to
the SPDE; see Section \ref{paperC:sec:galerkin} for details.

In this paper we extend the work by \citet{lindgren10} and
construct a
new flexible class of spatial models by considering a generalization of
\eqref{paperC:sde}. This model class contains a wide family of
covariance functions, including both the Mat\'{e}rn family and
oscillating covariance functions, and it maintains all desirable
properties of the Markov approximated Mat\'{e}rn model, such as
computational efficiency, easy nonstationary extensions and
applicability to data on general smooth manifolds.

The model class is introduced in Section \ref{paperC:sec:model}, with
derivations of some basic properties, examples of covariance functions
that can be obtained from these models and a discussion on
nonstationary extensions. Section \ref{paperC:sec:galerkin} gives a
review of the Hilbert space approximation technique and shows how it
can be extended to give computationally efficient representations also
for this new model class. In Section~\ref{paperC:sec:paramest} a
numerical parameter estimation procedure for the nested SPDE models is
presented, and the section concludes with a discussion on computational
complexity for parameter estimation and Kriging prediction.
In Section \ref{paperC:sec:ozone} the model class is used to analyze
the TOMS ozone data. In particular, all measurements available from
October 1st, 1988 in the spatially and temporally irregular ``Level 2''
version of the data set are used. This data set contains approximately
180,000 measurements, and the nonstationary version of the model
class is used to construct estimates of the ozone field for that
particular day. Finally, Section \ref{paperC:sec:conclusions} contains
some concluding remarks and suggestions for further work.

\section{Stationary nested SPDE models}\label{paperC:sec:model}
A limitation with the Mat\'{e}rn covariance family is that it does not
contain any covariance functions with negative values, such as
oscillating covariance functions. One way of constructing a larger
class of stochastic fields is to consider a generalization of the SPDE
\eqref{paperC:sde}:
%
\begin{equation}\label{paperC:eq:model}
\mathcal{L}_1 X(\mathbf{s}) = \mathcal{L}_2 \mathcal{W}(\mathbf{s}),
\end{equation}
for some linear operators $\mathcal{L}_1$ and
$\mathcal{L}_2$. If $\mathcal{L}_1$ and $\mathcal{L}_2$ are commutative
operators, \eqref{paperC:eq:model} is equivalent to the following
system of
nested SPDEs:
\begin{eqnarray}\label{paperC:eq:model2}
\mathcal{L}_1 X_0(\mathbf{s}) &=& \mathcal{W}(\mathbf{s}),\nonumber
\\[-8pt]\\[-8pt]
X(\mathbf{s}) &=& \mathcal{L}_2 X_0(\mathbf{s}).\nonumber
\end{eqnarray}
This representation gives us an interpretation of
the consequence of the additional differential operator
$\mathcal{L}_2$: $X(\mathbf{s})$ is simply $\mathcal{L}_2$ applied
to the
solution one would get to \eqref{paperC:eq:model} if $\mathcal{L}_2$
was the identity operator. Equation
\eqref{paperC:eq:model} generates a large class of random
fields, even if the operators $\mathcal{L}_1$ and $\mathcal{L}_2$ are
restricted to operators closely related to \eqref{paperC:sde}. One of
the simplest
extensions of the Mat\'{e}rn model is to let $\mathcal{L}_1$ be the
same as in \eqref{paperC:sde} and use $\mathcal{L}_2 = (b +
\mathbf{B}^{\top}\nabla)$, where $\nabla$ is the gradient, $b\in
\mathbb{R}
$, and
$\mathbf{B}\in\mathbb{R}^d$. The equation then is
%
\begin{equation}\label{paperC:eq:smallmodel}
(\kappa^2-\Delta)^{{\alpha/2}}X(\mathbf{s}) = (b + \mathbf
{B}^{\top
}\nabla)\mathcal{W}(\mathbf{s}),
\end{equation}
and $X(\mathbf{s})$ is a weighted sum of a Mat\'{e}rn field and its
directional derivative in the direction determined by the vector
$\mathbf{B}$. This model is closely related to the models introduced
in \citet{jun07} and \citet{jun08}, and the connection is discussed
later in Section \ref{paperC:sec:ozone}. To get a larger class of
models, the orders of the operators $\mathcal{L}_1$ and
$\mathcal{L}_2$ can be increased, and to get a class of stochastic
fields that is easy to work with, the operators are written as products,
where each factor in the product is equal to one of the operators in
\eqref{paperC:eq:smallmodel}. Thus, let
%
\begin{equation}\label{paperC:eq:L1}
\mathcal{L}_1 = \prod_{i=1}^{n_1} (\kappa_i^2-\Delta)^{
{\alpha_i/2}}
\end{equation}
for $\alpha_i \in\mathbb{N}$ and $\kappa_i^2>0$, and use
%
\begin{equation}\label{paperC:eq:L2}
\mathcal{L}_2 = \prod_{i=1}^{n_2} (b_i + \mathbf{B}_i^{\top}\nabla)
\end{equation}
for $b_i\in\mathbb{R}$ and $\mathbf{B}_i\in\mathbb{R}^d$. Hence,
the SPDE
generating the
class of nested SPDE models is
%
\begin{equation}\label{paperC:eq:model3}
\Biggl(\prod_{i=1}^{n_1} (\kappa^2-\Delta)^{{\alpha
_i/2}}\Biggr)X(\mathbf{s}) =
\Biggl(\prod_{i=1}^{n_2} (b_i + \mathbf{B}_i^{\top}\nabla)\Biggr)\mathcal{W}
(\mathbf{s}).
\end{equation}

There are several alternative equations one might consider; one
could, for example, let $\mathcal{L}_2$ be on the same form as
$\mathcal{L}_1$, or allow for anisotropic operators on the form
$(1-\nabla^{\top}\mathbf{A}\nabla)$ for some positive definite matrix
$\mathbf{A}$. However, to limit our scope, we will from now on only
consider model \eqref{paperC:eq:model3}.

\subsection{Properties in $\mathbb{R}^d$}

In this section some basic properties of random fields generated
by \eqref{paperC:eq:model3}, when $\mathbf{s}\in\mathbb{R}^d$, are given.
First note that all Mat\'{e}rn fields with shape parameters
satisfying $\nu+ d/2 \in\mathbb{N}$ are contained in the class of
stochastic fields generated by \eqref{paperC:eq:model3} since
$(\kappa^2-\Delta)^{{\alpha/2}}$ can be written on the form
\eqref{paperC:eq:L1} for these values of $\nu$. Also note that the
order of the operator $\mathcal{L}_2$ cannot be larger than the order
of $\mathcal{L}_1$ if $X(\mathbf{s})$ should be at least as ``well
behaved'' as
white noise; hence, one must have $\sum_{i=1}^{n_1}\alpha_i \geq n_2$.
The smoothness of $X(\mathbf{s})$ is determined by the difference of the
orders of the
operators $\mathcal{L}_1$ and $\mathcal{L}_2$. In order to make a precise
statement about this, the spectral density of $X(\mathbf{s})$ is needed.

\begin{prop}\label{paperC:specprop}
The spectral density for $X(\mathbf{s})$ defined by \eqref
{paperC:eq:model3} is given by
\begin{eqnarray*}
S(\mathbf{k}) =\frac{\phi^2}{(2\pi)^d} \frac{\prod_{j=1}^{n_2}
(b_j^2 +
\mathbf{k}^{\top}\mathbf{B}_j \mathbf{B}_j^{\top}\mathbf
{k})}{\prod_{j=1}^{n_1}
(\kappa_j^2+\|\mathbf{k}\|^2)^{\alpha_j}}.
\end{eqnarray*}
\end{prop}

This proposition is easily proved using linear filtering theory [see,
for example, \citet{yaglom87}]. Given the spectral density of
$X(\mathbf{s})$, the following proposition regarding the sample function
regularity can be proved using Theorem~3.4.3 in \citet{adler81}.

\begin{prop}\label{paperC:prop:cont}
$X(\mathbf{s})$ defined by \eqref{paperC:eq:model3} has almost surely
continuous sample
functions if $2\sum_{i=1}^{n_1}\alpha_i-2n_2 > d$.
\end{prop}

Because the stochastic field $X(\mathbf{s})$ is generated by the SPDE
\eqref{paperC:eq:model3}, the following corollary regarding sample
path differentiability is also easily proved using the fact that the
directional derivative of $X(\mathbf{s})$ is in the class of nested
SPDE models.

\begin{cor}\label{paperC:prop:corr}
Given that $2\sum_{i=1}^{n_1}\alpha_i-2n_2 - d>m$, the $m$th order
directional derivative of $X(\mathbf{s})$, $(\mathbf{B}^{\top}\nabla)^m
X(\mathbf{s})$, has almost surely continuous sample functions.
\end{cor}

Hence, as $2\sum_{i=1}^{n_1}\alpha_i-2n_2$ increases, the sample
paths become smoother, and eventually become
differentiable, twice differentiable, and so on. One could also
give a more precise characterization of the sample path regularity
using the notion of H\"{o}lder continuity. This is (more or less)
straightforward using properties of index-$\beta$ random fields
[\citet{adler81}], but outside the scope of this article.

A closed-form expression for the covariance function is not that
interesting since none of the methods that are later presented for
parameter estimation, spatial prediction or model validation require an
expression for the covariance function; however, if one were to use
some technique that requires the covariance function, it can be
derived. An expression for the general case is quite complicated,
and will not be presented here. Instead we present a recipe for
calculating the covariance function for given parameters of the SPDE,
with explicit results for a few examples.

\begin{figure}[t]

\includegraphics{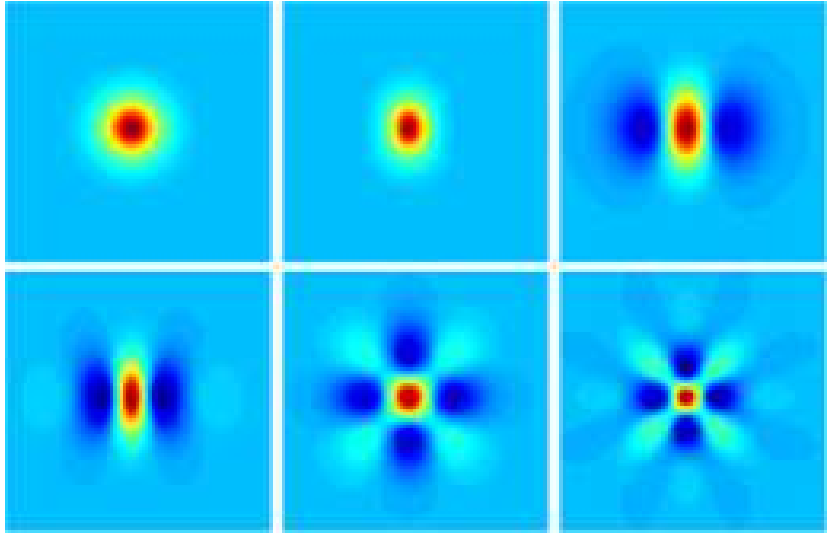}

\caption{Covariance functions of random fields obtained from model
\protect\eqref{paperC:eq:model3} with parameters from Example~\protect\ref{ex1} (top left),
Example~\protect\ref{ex2} (top middle and right), Example~\protect\ref{ex3} (bottom left and middle)
and Example~\protect\ref{ex4} (bottom right).}
\label{paperC:fig:covs}
\end{figure}

\begin{figure}[t]

\includegraphics{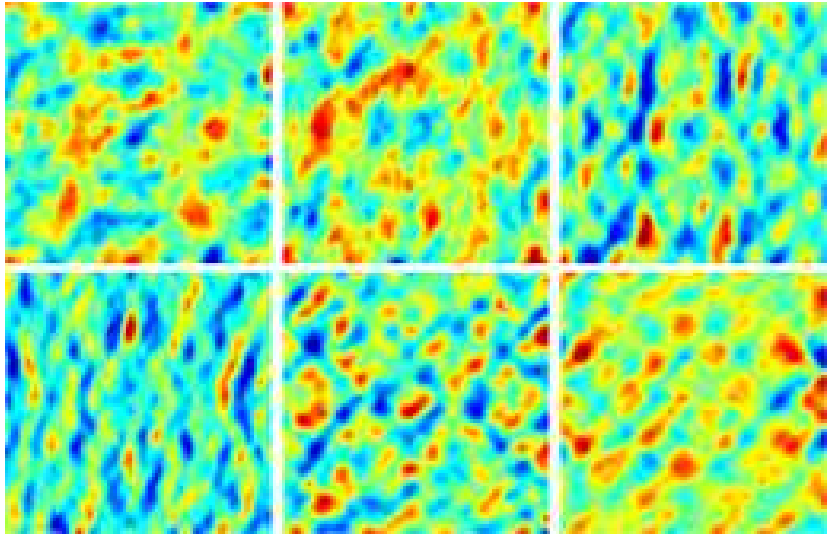}

\caption{Realizations of random fields obtained from model
\protect\eqref{paperC:eq:model3} with different parameters. The realization in
each panel corresponds to a stochastic field with the covariance
function shown in the corresponding panel in Figure~\protect\ref{paperC:fig:covs}.}
\label{paperC:fig:realizations}
\end{figure}

To calculate the covariance function of $X(\mathbf{s})$, first calculate
the covariance function, $C_{X_0}(\mathbf{h})$, of $X_0(\mathbf{s})$,
given by
\eqref{paperC:eq:model2}. Given this covariance function, the covariance
function for $X(\mathbf{s})$ is obtained as
\begin{eqnarray*}
C(\mathbf{h}) = \Biggl(\prod_{i=1}^{n_2} (b_i - \nabla^{\top}\mathbf
{B}_i\mathbf{B}_i^{\top}\nabla)\Biggr)C_{X_0}(\mathbf{h}).
\end{eqnarray*}
The motivation for this expression is again directly from linear
filter theory, and the $d$-dimensional equivalent of the formula for
the covariance function for a differentiated stochastic process,
$r_{X'}(\tau) = - r_{X}''(\tau)$. To get an expression for
$C_{X_0}(\mathbf{h})$, first use Proposition \ref{paperC:specprop} with
$\mathcal{L}_2=I$ to get the spectral density of $X_0(\mathbf{s})$.
Using a
partial fraction decomposition, the spectral density can be written as
%
\begin{equation}\label{paperC:eq:spec}
S_{X_0}(\mathbf{k}) =\frac{\phi^2}{(2\pi)^d}\sum_{i=1}^n\sum
_{j=1}^{\alpha
_i}\frac{p_{i,j}}{(\kappa_i^2+\|\mathbf{k}\|^2)^j},
\end{equation}
where $p_{i,j}$ is a real constant which can be found using the
Heaviside cover-up method [see, for example, \citet{thomas95}, page
523]. Now, by taking the inverse Fourier
transform of \eqref{paperC:eq:spec}, the covariance function for
$X_0(\mathbf{s})$ is
\begin{eqnarray*}
C_{X_0}(\mathbf{h}) =\sum_{i=1}^n\sum_{j=1}^{\alpha_i}
p_{i,j}C_{\kappa
_i}^{j}(\mathbf{h}),
\end{eqnarray*}
where $C_{\kappa}^{\nu}(\mathbf{h})$ denotes a Mat\'{e}rn covariance function
with shape parameter $\nu$, scale parameter $\kappa$ and variance
parameter $\phi^2$. The final step is to use that the derivative of a
Mat\'{e}rn
covariance function can be expressed using a Mat\'{e}rn covariance
with another shape parameter. More precisely, one has
\begin{eqnarray*}
\frac{\partial}{\partial h_i}C_{\kappa}^{\nu}(\mathbf{h}) = -\frac
{h_i}{2\nu
}C_{\kappa}^{\nu-1}(\mathbf{h}),
\end{eqnarray*}
where $h_i$ denotes the $i$th component of the vector $\mathbf{h}$. Using
these calculations, one can obtain the covariance function for any
field given by \eqref{paperC:eq:model3}. We conclude this section by showing
the covariance function for some simple cases in $\mathbb{R}^2$. The
covariance functions for these examples are shown in
Figure~\ref{paperC:fig:covs}, and realizations of Gaussian
processes with
these covariance functions are shown in Figure~\ref{paperC:fig:realizations}.

\begin{example}\label{ex1}
With $\mathcal{L}_1 = (\kappa^2-\Delta)^{{\alpha/2}}$ and
$\mathcal{L}_2$
as the identity operator, the standard Mat\'{e}rn covariance
function \eqref{paperC:eq:matern} is obtained, shown in the top left panel
of Figure \ref{paperC:fig:covs}.
\end{example}

\begin{example}\label{ex2}
The simplest nested SPDE model \eqref{paperC:eq:smallmodel} has the covariance
function
\begin{eqnarray*}
C(\mathbf{h}) = bC_{\kappa}^{\nu}(\mathbf{h}) +
\frac{\mathbf{B}^{\top}\mathbf{B}}{2\nu}C_{\kappa}^{\nu
-1}(\mathbf{h}) -
\frac{\mathbf{h}^{\top}\mathbf{B}\mathbf{B}^{\top}\mathbf
{h}}{4\nu(\nu-1)}C_{\kappa
}^{\nu
-2}(\mathbf{h}).
\end{eqnarray*}
A stochastic field with this covariance function is obtained as a
weighted sum of a Mat\'{e}rn field $X_0(\mathbf{s})$ and its directional
derivative in the direction of $\mathbf{B}$. The field therefore has a
Mat\'{e}rn-like behavior in the direction perpendicular to $\mathbf{B}$
and an oscillating behavior in the direction of $\mathbf{B}$. In the upper
middle panel of Figure \ref{paperC:fig:covs}, this covariance function
is shown for $\mathbf{B} = (1,0)^{\top}$, $\nu=3$, and $b = 5$.
In the upper right panel of Figure \ref{paperC:fig:covs}, it is shown
for $\mathbf{B} = (1,0)^{\top}$, $\nu=3$, and $b = 0$.
\end{example}

\begin{example}\label{ex3}
The number of zero crossings of the covariance function in the
direction of $\mathbf{B}$ is at most $n_2$. In the previous example
we had $n_2 = 1$, and to obtain a more oscillating covariance function, the
order of $\mathcal{L}_2$ can be increased by one:
\begin{eqnarray*}
(\kappa^2-\Delta)^{{\alpha/2}}X(\mathbf{s}) = (b_1 + \mathbf
{B}_1^{\top
}\nabla) (b_2 + \mathbf{B}_2^{\top}\nabla)\mathcal{W}(\mathbf{s}).
\end{eqnarray*}
This model has the covariance function
\begin{eqnarray*}
C(\mathbf{h}) &=& b_1b_2C_{\kappa}^{\nu}(\mathbf{h}) +
\frac{b_2\mathbf{B}_1^{\top}\mathbf{B}_1+b_1\mathbf{B}_2^{\top
}\mathbf{B}_2}{2\nu
}C_{\kappa}^{\nu-1}(\mathbf{h})
\\
&&{}+ \frac{2(\mathbf{B}_2^{\top}\mathbf{B}_1)^2 + \mathbf{B}_1^{\top
}\mathbf{B}_1\mathbf{B}_2^{\top}\mathbf{B}_2 - \mathbf{h}^{\top
}(b_1\mathbf{B}_2\mathbf{B}_2^{\top
}+b_2\mathbf{B}_1\mathbf{B}_1^{\top})\mathbf{h}}{2^2\nu(\nu
-1)}C_{\kappa
}^{\nu
-2}(\mathbf{h})
\\
&&{}-\frac{\mathbf{h}^{\top}(\mathbf{B}_1\mathbf{B}_2^{\top}\mathbf
{B}_2\mathbf{B}_1^{\top
}+4\mathbf{B}_1\mathbf{B}_1^{\top}\mathbf{B}_2\mathbf{B}_2^{\top
}+\mathbf{B}_2\mathbf{B}_1^{\top}\mathbf{B}_1\mathbf{B}_2^{\top
})\mathbf{h}}{2^3\nu(\nu-1)(\nu
-2)}C_{\kappa}^{\nu-3}(\mathbf{h})
\\
&&{}+ \frac{(\mathbf{B}_1^{\top}\mathbf{h}\mathbf{h}^{\top}\mathbf
{B}_2)^2}{2^4\nu(\nu
-1)(\nu-2)(\nu-3)}C_{\kappa}^{\nu-4}(\mathbf{h}).
\end{eqnarray*}
In the bottom left panel of Figure \ref{paperC:fig:covs} this covariance
function is shown for $\nu=5$, $b_1=b_2=0$ and $\mathbf{B}_1 =
\mathbf{B}_2 =
(1,0)^{\top}$. With these parameters, the covariance function is
similar to the covariance function in the previous example, but with
one more zero crossing in the direction of $\mathbf{B}$. For this specific
choice of parameters, the expression for the covariance function can
be simplified to
\begin{eqnarray*}
C(\mathbf{h}) = 3\gamma_2 C_{\kappa}^{\nu-2}(\mathbf{h}) - 6\gamma
_3 h_1^2
C_{\kappa}^{\nu-3}(\mathbf{h})+ \gamma_4 h_1^4C_{\kappa}^{\nu
-4}(\mathbf{h}),
\end{eqnarray*}
where $\gamma_k = (2^k \Pi_{i=0}^{k-1}(\nu-k))^{-1}$. In the bottom
middle panel of Figure \ref{paperC:fig:covs} the covariance function is
shown for
$\nu=5$, $b_1=b_2=0$, $\mathbf{B}_1 = (1,0)^{\top}$, and $\mathbf
{B}_2 =
(0,1)^{\top}$. Thus, the field $X_0(\mathbf{s})$ is differentiated in two
different directions, and the covariance function for $X(\mathbf{s})$
therefore is oscillating in two directions. For these parameters, the
covariance function can be written as
\begin{eqnarray*}
C(\mathbf{h}) = \gamma_2 C_{\kappa}^{\nu-2}(\mathbf{h})-\gamma_3
\mathbf{h}^{\top
}\mathbf{h}C_{\kappa}^{\nu-3}(\mathbf{h})+ \gamma_4h_1h_2C_{\kappa
}^{\nu
-4}(\mathbf{h}).
\end{eqnarray*}
\end{example}

\begin{example}\label{ex4}
The bottom right panel of Figure \ref{paperC:fig:covs} shows a covariance
function for the nested SPDE
\begin{eqnarray*}
(\kappa^2-\Delta)^{{\alpha/2}}X(\mathbf{s}) = (b_1 + \mathbf
{B}_1^{\top
}\nabla)^2 (b_2 + \mathbf{B}_2^{\top}\nabla)^2\mathcal{W}(\mathbf{s}).
\end{eqnarray*}
As in the previous examples, the covariance function for a stochastic
field generated by this SPDE can be calculated and written on the form
\begin{eqnarray*}
C(\mathbf{h}) = \sum_{k=0}^{8}\gamma_k f_k(\mathbf{h}) C_{\kappa
}^{\nu
-k}(\mathbf{h}),
\end{eqnarray*}
where $f_k(\mathbf{h}), k=0,\ldots,8,$ are functions depending on
$\mathbf{h}$
and the
parameters in the SPDE. Without any restrictions on the parameters, it
is a
rather tedious exercise to calculate the functions $f_k(\mathbf{h})$, and
we therefore only show them for the specific set of parameters that
are used in Figure \ref{paperC:fig:covs}: $\nu=7$, $b_1=b_2=0$,
$\mathbf{B}_1 =
(1,0)^{\top}$ and $\mathbf{B}_2 = (0,1)^{\top}$. In this case
$f_0(\mathbf{h})=f_1(\mathbf{h})=f_2(\mathbf{h})=0$, and the
covariance function is
\begin{eqnarray*}
C(\mathbf{h}) &=& 9\gamma_4 C_{\kappa}^{\nu-4}(\mathbf{h}) -
18\gamma_5
\mathbf{h}^{\top}\mathbf{h}
C_{\kappa}^{\nu-5}(\mathbf{h}) +
3\gamma_6(h_1^4+h_2^4+12h_1^2h_2^2)C_{\kappa}^{\nu-6}(\mathbf{h})
\\
&&{}- 6\gamma_7 h_1^2h_2^2 \mathbf{h}^{\top}\mathbf{h} C_{\kappa
}^{\nu
-7}(\mathbf{h})
+ \gamma_8 h_1^4 h_2^4 C_{\kappa}^{\nu-8}(\mathbf{h}).
\end{eqnarray*}
\end{example}

\subsection{Nonstationary nested SPDE models}\label{paperC:sec:nonstat}
Nonstationarity can be introduced in the nested SPDE models by allowing
the parameters $\kappa_i$, $b_i$ and $\mathbf{B}_i$ to be spatially varying:
\begin{eqnarray}\label{paperC:eq:modelnonstat}
\Biggl(\prod_{i=1}^{n_1} \bigl(\kappa_i^2(\mathbf{s})-\Delta\bigr)^{{\alpha
_i/2}}\Biggr)X_0(\mathbf{s}) &=& \mathcal{W}(\mathbf{s}),\nonumber
\\[-8pt]\\[-8pt]
X(\mathbf{s}) &=& \Biggl(\prod_{i=1}^{n_2} \bigl(b_i(\mathbf{s}) + \mathbf
{B}_i(\mathbf{s})^{\top
}\nabla\bigr)\Biggr)X_0(\mathbf{s}).\nonumber
\end{eqnarray}
If the parameters are spatially varying, the two operators are no
longer commutative, and the solution to \eqref{paperC:eq:modelnonstat}
is not necessarily equal to the solution of
%
\begin{equation}\label{paperC:eq:modelnonstatshort}
\Biggl(\prod_{i=1}^{n_1} \bigl(\kappa_i^2(\mathbf{s})-\Delta\bigr)^{{\alpha
_i/2}}\Biggr)X(\mathbf{s}) = \Biggl(\prod_{i=1}^{n_2} \bigl(b_i(\mathbf{s}) +
\mathbf{B}_i(\mathbf{s})^{\top}\nabla\bigr)\Biggr)\mathcal{W}(\mathbf{s}).
\end{equation}
For nonstationary models, we will from now on only study the system of
nested SPDEs \eqref{paperC:eq:modelnonstat}, though it should be noted
that the methods presented in the next sections can be applied to
\eqref
{paperC:eq:modelnonstatshort} as well.

One could potentially use an approach where the spatially varying
parameters also are modeled as stochastic fields, but to be able to
estimate the
parameters efficiently, it is easier to assume that each
parameter can be written as a weighted sum of some known regression
functions. In Section~\ref{paperC:sec:ozone} this approach is used
for a nested SPDE model on the sphere. In this case, one needs a regression
basis $\{\psi_j(\mathbf{s})\}$ for the vector fields $\mathbf
{B}_i(\mathbf{s})$ on
the sphere. Explicit expressions for such a basis are given in the
\hyperref[paperC:sec:vectorbasis]{Appendix}.

\section{Computationally efficient representations}\label{paperC:sec:galerkin}
In the previous section covariance functions for some
examples of nested SPDE models were derived. Given the covariance
function, standard spatial statistics techniques can be used for
parameter estimation, spatial prediction and model simulation. Many of these
techniques are, however, computationally infeasible for large data
sets. Thus, in order to use the model for large
environmental data sets, such as the ozone data studied in Section
\ref{paperC:sec:ozone}, a more computationally efficient
representation of the model class is needed. In this section the
Hilbert space approximation technique by \citet{lindgren10} is used to derive
such a representation.

The key idea in \citet{lindgren10} is to approximate the solution to
the SPDE
$\mathcal{L}_1X_0(\mathbf{s}) = \mathcal{W}(\mathbf{s})$ in some
approximation space
spanned by basis functions $\varphi_1(\mathbf{s}), \ldots,
\varphi_n(\mathbf{s})$. The method is most efficient if these basis
functions have compact support, so, from now on, it is assumed that
$\{\varphi_i\}$ are local basis functions. The weak solution of the
SPDE with
respect to the approximation space can be written as
$\tilde{x}(\mathbf{s}) = \sum_{i=1}^n w_i \varphi_i(\mathbf{s})$,
where the
stochastic weights $\{w_i\}_{i=1}^n$ are chosen such that the weak
formulation of the SPDE is satisfied:
%
\begin{equation}\label{paperC:eq:weak}
[\langle{\varphi_i},{\mathcal{L}_1\tilde{x}} \rangle_{\Omega
}]_{i=1,\ldots, n}
\stackrel{D}{=} [\langle{\varphi_i},{\mathcal{W}} \rangle_{\Omega
}]_{i=1,\ldots, n}.
\end{equation}
Here $\stackrel{D}{=}$ denotes equality in distribution, $\Omega$ is
the manifold on which $\mathbf{s}$ is defined, and $\langle{f},{g}
\rangle_{\Omega} =
\int_{\Omega}f(\mathbf{s})g(\mathbf{s})\,\mathrm{d}\mathbf{s}$ is
the scalar product on
$\Omega$.
As an illustrative example, consider the first fundamental case
$\mathcal{L}_1 = \kappa^2 - \Delta$. One~has
\begin{eqnarray*}
\langle{\varphi_i},{\mathcal{L}_1\tilde{x}} \rangle_{\Omega} =
\sum_{j=1}^n
w_j\langle{\varphi_i},{\mathcal{L}_1\varphi_j} \rangle_{\Omega},
\end{eqnarray*}
so by introducing a matrix $\mathbf{K}$, with elements $\mathbf
{K}_{i,j} =
\langle{\varphi_i},{\mathcal{L}_1\varphi_j} \rangle_{\Omega}$,
and the vector
$\mathbf{w} = (w_1, \ldots, w_n)^{\top}$, the left-hand side of
\eqref{paperC:eq:weak} can be written as $\mathbf{K}\mathbf{w}$. Since
\begin{eqnarray*}
\langle{\varphi_i},{\mathcal{L}_1\varphi_j} \rangle_{\Omega} &=&
\kappa^2\langle{\varphi_i},{\varphi_j} \rangle_{\Omega}-\langle
{\varphi_i},{\Delta\varphi_j} \rangle_{\Omega}
\\
&=&\kappa^2\langle{\varphi_i},{\varphi_j} \rangle_{\Omega}+\langle
{\nabla\varphi_i},{\nabla\varphi_j} \rangle_{\Omega},
\end{eqnarray*}
the matrix $\mathbf{K}$ can be written as $\mathbf{K} = \kappa
^2\mathbf{C} +
\mathbf{G}$, where $\mathbf{C}_{i,j} = \langle{\varphi_i},{\varphi
_j} \rangle_{\Omega}$
and $\mathbf{G}_{i,j} = \langle{\nabla\varphi_i},{\nabla\varphi
_j} \rangle_{\Omega}$.
The right-hand side of \eqref{paperC:eq:weak} can be shown to be Gaussian
with mean zero and covariance $\mathbf{C}$.
For the Hilbert space approximations, it is natural to work with
the canonical representation, $\mathbf{x}\sim\mathsf{N}_{C}(\mathbf
{b},\mathbf{Q})$, of
the Gaussian distribution. Here, the precision matrix $\mathbf{Q}$ is the
inverse of the covariance matrix, and the vector $\mathbf{b}$ is connected
to the mean, $\bolds{\mu}$, of the Gaussian distribution through the
relation $\bolds{\mu} =
\mathbf{Q}^{-1}\mathbf{b}$. Thus, if $\mathbf{K}$ is
invertible, one has
\begin{eqnarray*}
\mathbf{K}\mathbf{w} \sim\mathsf{N}_C(\mathbf{0}, \mathbf
{C}^{-1}) \quad\Longleftrightarrow\quad \mathbf{w}
\sim\mathsf{N}_C(\mathbf{0},\mathbf{K}\mathbf{C}^{-1}\mathbf{K}).
\end{eqnarray*}
For the second fundamental case, $\mathcal{L}_1 = (\kappa^2
- \Delta)^{1/2}$, \citet{lindgren10} show that $\mathbf{w} \sim
\mathsf{N}
_C(\mathbf{0},\mathbf{K})$.
Given these two fundamental cases, the weak
solution to $\mathcal{L}_1X_0(\mathbf{s}) = \mathcal{W}(\mathbf{s})$,
for any
operator on the form \eqref{paperC:eq:L1}, can be obtained
recursively. If, for example, $\mathcal{L}_1 = (\kappa^2 - \Delta)^2$,
the solution is obtained by solving $(\kappa^2 - \Delta)X_0(\mathbf
{s}) =
\tilde{x}(\mathbf{s})$, where $\tilde{x}$ is the weak solution to the
first fundamental case.

The iterative way of constructing solutions can be extended to calculate
weak solutions to \eqref{paperC:eq:model3} as well. Let $\tilde{x}_0 =
\sum_{i=1}^n w_i^0 \varphi_i(\mathbf{s})$ be a weak solution to
$\mathcal{L}_1X_0(\mathbf{s}) = \mathcal{W}(\mathbf{s})$, and let
$\mathbf{Q}_{X_0}$
denote the precision for the weights $\mathbf{w}_0=(w_1^0, \ldots,
w_n^0)^{\top}$. Substituting $X_0$ with $\tilde{x}_0$ in the second
equation of \eqref{paperC:eq:model}, the weak formulation of the
equation is
\begin{eqnarray}\label{paperC:eq:weak2}
[\langle{\varphi_i},{\tilde{x}} \rangle_{\Omega}]_{i=1,\ldots, n}
&\stackrel{D}{=}&
[\langle{\varphi_i},{\mathcal{L}_2\tilde{x}_0} \rangle_{\Omega
}]_{i=1,\ldots, n}\nonumber
\\[-8pt]\\[-8pt]
&=& \Biggl[\sum_{j=1}^n w_j^0 \langle{\varphi_i},{\mathcal{L}_2 \varphi
_j} \rangle_{\Omega}\Biggr]_{i=1,\ldots, n}.\nonumber
\end{eqnarray}
First consider the case of an order-one operator $\mathcal{L}_2
= b_1 + \mathbf{B}_1^{\top}\nabla$. By introducing the matrix
$\mathbf{H}_1$ with
elements $\mathbf{H}_{1 i,j} =
\langle{\varphi_i},{\mathcal{L}_2\varphi_j} \rangle_{\Omega}$,
the right-hand side
of \eqref{paperC:eq:weak2} can be written as $\mathbf{H}_1\mathbf
{w}_0$. Introducing
the vector $\mathbf{w} = (w_1, \ldots, w_n)^{\top}$, the left-hand side
of \eqref{paperC:eq:weak2} can be written as $\mathbf{C}\mathbf{w}$,
and one has
\begin{eqnarray*}
\mathbf{w} = \mathbf{C}^{-1}\mathbf{H}_1\mathbf{w}_0\quad \Longrightarrow\quad
\mathbf{w} \sim\mathsf{N}
_{C}(\mathbf{0}, \mathbf{C}\mathbf{H}_1^{-\top}\mathbf
{Q}_{X_0}\mathbf{H}_1^{-1}\mathbf{C}).
\end{eqnarray*}
Now, if $\mathcal{L}_2$ is on the form \eqref{paperC:eq:L2}, the
procedure can be used recursively, in the same way as when producing
higher order Mat\'{e}rn fields. For example,~if
\begin{eqnarray*}
\mathcal{L}_2= (b_1 + \mathbf{B}_1^{\top}\nabla)(b_2 + \mathbf
{B}_2^{\top
}\nabla),
\end{eqnarray*}
the solution is obtained by solving $X(\mathbf{s}) = (b_2 +
\mathbf{B}_2^{\top}\nabla)\tilde{x}(\mathbf{s})$, where $\tilde
{x}$ is the
weak solution to the previous example. Thus, when $\mathcal{L}_2$ is on
the form \eqref{paperC:eq:L2}, one has
\begin{eqnarray*}
\mathbf{w} \sim\mathsf{N}_{C}(\mathbf{0},\mathbf{H}^{-\top
}\mathbf{Q}_{X_0}\mathbf{H}^{-1}),\qquad
\mathbf{H} = \mathbf{C}^{-1}\mathbf{H}_{n_2}\mathbf{C}^{-1}\mathbf
{H}_{n_2-1}\cdots
\mathbf{C}^{-1}\mathbf{H}_{1},
\end{eqnarray*}
where each factor $\mathbf{H}_i$ corresponds to the $\mathbf{H}$-matrix
obtained in the $i$th step in the recursion.

\subsection{Nonstationary fields}
As mentioned in \citet{lindgren10}, the Hilbert space approximation
technique can also be used for nonstationary models, and the technique
extends to the nested SPDE models as well. One again begins by finding
the weak solution of the first part of the system, $\mathcal
{L}_1(\mathbf{s})X_0(\mathbf{s}) = \mathcal{W}(\mathbf{s})$. The
iterative procedure is used for
obtaining approximations of high-order operators, so the fundamental
step is to find the weak solution to the equation when $\mathcal{L}_1 =
(\kappa^2(\mathbf{s}) - \Delta)$. Consider the weak formulation\vspace*{2pt}
%
\begin{equation}\label{paperC:eq:weak3}
\bigl[\bigl\langle{\varphi_i},{\bigl(\kappa^2(\mathbf{s}) - \Delta\bigr)\tilde{x}}
\bigr\rangle_{\Omega
}\bigr]_{i=1,\ldots, n} \stackrel{D}{=} [\langle{\varphi_i},{\mathcal{W}}
\rangle_{\Omega
}]_{i=1,\ldots, n},\vspace*{2pt}
\end{equation}
and note that the right-hand side of the equation is the
same as in the stationary case, $\mathsf{N}_{C}(\mathbf{0},\mathbf
{C}^{-1})$. Now,
using that\vspace*{2pt}
\begin{eqnarray*}
\bigl\langle{\varphi_i},{\bigl(\kappa^2(\mathbf{s}) - \Delta\bigr)\tilde{x}}
\bigr\rangle_{\Omega} &=&
\langle{\varphi_i},{\kappa^2(\mathbf{s})\tilde{x}} \rangle
_{\Omega} -
\langle{\varphi_i},{\Delta\tilde{x}} \rangle_{\Omega}
\\[2pt]
&=& \langle{\varphi_i},{\kappa^2(\mathbf{s})\tilde{x}} \rangle
_{\Omega} +
\langle{\nabla\varphi_i},{\nabla\tilde{x}} \rangle_{\Omega},\vspace*{2pt}
\end{eqnarray*}
the left-hand side of \eqref{paperC:eq:weak3} can be written as
$(\tilde{\mathbf{C}} + \mathbf{G})\mathbf{w}_0$, where $\mathbf
{G}$ and $\mathbf{w}_0$
are the same as in the stationary case and $\tilde{\mathbf{C}}$ is a
matrix with elements\vspace*{2pt}
\begin{eqnarray}\label{paperC:eq:Capprox}
\tilde{\mathbf{C}}_{i,j} &=& \langle{\varphi_i},{\kappa^2(\mathbf
{s})\varphi_j} \rangle_{\Omega} =
\int_{\Omega}\kappa^2(\mathbf{s})\varphi_i(\mathbf{s})\varphi
_j(\mathbf{s})\,\mathrm{d}\mathbf{s}\nonumber
\\[-7pt]\\[-7pt]
&\approx&\kappa^2(\mathbf{s}_j)\int_{\Omega}\varphi_i(\mathbf
{s})\varphi
_j(\mathbf{s})\,\mathrm{d}\mathbf{s} = \kappa^2(\mathbf{s}_j)\mathbf
{C}_{i,j}.\nonumber\vspace*{2pt}
\end{eqnarray}
Since $\{\varphi_i\}$ is assumed to be a local basis, such as
B-spline wavelets or some other functions with compact support, the
locations $\mathbf{s}_j$ can, for example, be chosen as the centers of the
basis functions $\varphi_j(\mathbf{s})$. The error in the
approximation of
$\tilde{\mathbf{C}}$ is then small if $\kappa^2(\mathbf{s})$
varies slowly compared to the spacing of the basis functions
$\varphi_j$. From equation \eqref{paperC:eq:Capprox}, one has
$\tilde{\mathbf{C}}~=~\mathbf{C}\bolds{\kappa}$, where $\bolds
{\kappa}$ is a
diagonal matrix with elements $\bolds{\kappa}_{j,j} = \kappa
^2(\mathbf{s}_j)$.
Finally, with $\mathbf{K} = \bolds{\kappa}\mathbf{C} + \mathbf
{G}$, one has\vspace*{2pt}
\begin{eqnarray*}
\mathbf{K}\mathbf{w}_0 \sim\mathsf{N}_{C}(\mathbf{0}, \mathbf
{C}^{-1}) \quad\Longrightarrow\quad\mathbf{w}_0
\sim\mathsf{N}_{C}(\mathbf{0},\mathbf{K}\mathbf{C}^{-1}\mathbf{K}).\vspace*{2pt}
\end{eqnarray*}
Now given the weak solution, $\tilde{x}_0$, to
$\mathcal{L}_1(\mathbf{s})X_0(\mathbf{s}) = \mathcal{W}(\mathbf
{s})$, substitute
$X_0$ with $\tilde{x}_0$ in the second equation of
\eqref{paperC:eq:model2} and consider the weak formulation of the
equation.
Since the solution to the full operator again can be found recursively,
only the fundamental case $\mathcal{L}_2 = b(\mathbf{s}) + \mathbf
{B}(\mathbf{s})^{\top}\nabla$ is considered. The weak formulation is
the same as
\eqref{paperC:eq:weak2}, and one has\vspace*{2pt}
\begin{eqnarray*}
\langle{\varphi_i},{\tilde{x}} \rangle_{\Omega}
&\stackrel{D}{=}&\langle{\varphi_i},{\mathcal{L}_2\tilde{x}_0}
\rangle_{\Omega}
= \bigl\langle{\varphi_i},{\bigl(b(\mathbf{s})+\mathbf{B}(\mathbf{s})^{\top
}\nabla\bigr)\tilde{x}_0} \bigr\rangle_{\Omega}
\\[2pt]
&=&\langle{\varphi_i},{b(\mathbf{s})\tilde{x}_0} \rangle_{\Omega}
+\langle{\varphi_i},{\mathbf{B}(\mathbf{s})^{\top}\nabla\tilde
{x}_0} \rangle_{\Omega}.\vadjust{\goodbreak}
\end{eqnarray*}

\noindent
Thus, the right-hand side of \eqref{paperC:eq:weak2} can be written as
$(\hat{\mathbf{C}} + \hat{\mathbf{H}})\mathbf{w}_0$, where
\begin{eqnarray*}
\hat{\mathbf{C}}_{i,j} &=& \langle{\varphi_i},{b(\mathbf
{s})\varphi_j} \rangle_{\Omega} =
\int_{\Omega}b(\mathbf{s})\varphi_i(\mathbf{s})\varphi_j(\mathbf
{s})\,\mathrm{d}\mathbf{s}
\approx b(\mathbf{s}_j)\mathbf{C}_{i,j},
\\
\hat{\mathbf{H}}_{i,j} &=&
\langle{\varphi_i},{\mathbf{B}(\mathbf{s})^{\top}\nabla\varphi
_j} \rangle_{\Omega
} =
\int_{\Omega}\varphi_i(\mathbf{s})\mathbf{B}(\mathbf{s})^{\top
}\nabla
\varphi_j(\mathbf{s})\,\mathrm{d}\mathbf{s}
\\
&\approx&\mathbf{B}(\tilde{\mathbf{s}}_j)^{\top}\int_{\Omega
}\varphi
_i(\mathbf{s})\nabla\varphi_j(\mathbf{s})\,\mathrm{d}\mathbf{s}.
\end{eqnarray*}
Here, similar approximations as in equation \eqref{paperC:eq:Capprox}
are used, so the expressions are accurate if the coefficients vary
slowly compared to the spacing of the basis functions
$\varphi_j$. The left-hand side of \eqref{paperC:eq:weak2} can again
be written as $\mathbf{C}\mathbf{w}$, so with $\mathbf{H}_1 = \hat
{\mathbf{C}} +
\hat{\mathbf{H}}$, one has $\mathbf{w}\sim\mathsf{N}_{C}(\mathbf{0},
\mathbf{C}\mathbf{H}_1^{-\top}\mathbf{Q}_{X_0}\mathbf
{H}_1^{-1}\mathbf{C})$.

\subsection{Practical considerations}\label{paperC:sec:explicit}
The integrals that must be calculated to get explicit expressions
for the matrices $\mathbf{C}$, $\mathbf{G}$ and $\mathbf{H}$ are
\begin{eqnarray*}
\int_{\Omega}\varphi_i(\mathbf{s})\varphi_j(\mathbf{s})\,\mathrm
{d}\mathbf{s},\qquad
\int_{\Omega}(\nabla\varphi_i(\mathbf{s}))^{\top}\nabla\varphi
_j(\mathbf{s})\,\mathrm{d}\mathbf{s} \quad\mbox{and} \quad\int_{\Omega
}\varphi_i(\mathbf{s})\nabla
\varphi_j(\mathbf{s})\,\mathrm{d}\mathbf{s}.
\end{eqnarray*}
In Section \ref{paperC:sec:ozone} a basis of piecewise linear
functions induced by a triangulation of the Earth is used; see Figure
\ref{paperC:fig:triangulation}. In this case, $\varphi_i(\mathbf
{s})$ is
a linear
function on each triangle, and $\nabla\varphi_i(\mathbf{s})$ is constant
on each triangle. The integrals, therefore, have simple
analytic expressions in this case, and more generally for all
piecewise linear bases induced by triangulated 2-manifolds.

Bases induced by triangulations have many desirable properties, such as
the simple analytic expression for the integrals and compact support.
They are, however, not orthogonal, which causes $\mathbf{C}^{-1}$ to be
dense. The weights $\mathbf{w}$, therefore, have a dense precision matrix,
unless $\mathbf{C}^{-1}$ is approximated with some sparse matrix. This
issue is addressed in \citet{lindgren10} by lowering the integration
order of
$\langle{\varphi_i},{\varphi_j} \rangle$, which results in an approximate,
diagonal $\mathbf{C}$ matrix, $\bar{\mathbf{C}}$, with diagonal
elements $\bar{\mathbf{C}}_{ii} = \sum_{k=1}^n\mathbf{C}_{ik}$.
\citet{bolin09}
perform numerical studies on how this approximation affects the
resulting covariance function of the process, and it is shown that the
error is small if the
approximation is used for piecewise linear bases. We will, therefore,
from now on use the approximate $\mathbf{C}$ matrix in all places where
$\mathbf{C}$ is used.

A natural question is how many basis functions one should use in order
to get a good approximation of the solution. The answer will depend on
the chosen basis, and, more importantly, on the specific parameters of
the SPDE model. \citet{bolin09} study the approximation error in the
Mat\'{e}rn case in $\mathbb{R}$ and $\mathbb{R}^2$ for different
bases, and in this
case the spacing of the basis functions compared to the range of the
covariance function for $X(\mathbf{s})$ determines the approximation error:
For a process with long range, fewer basis functions have to be used
than for a process with short range to obtain the same approximation
error. For more complicated, possibly nonstationary, nested SPDE
models, there is no easy answer to how the number of basis functions
should be chosen. Increasing the number of basis functions will
decrease the approximation error but increase the computational
complexity for the approximate model, so there is a trade-off between
accuracy and computational cost.
However, as long as the parameters vary slowly compared to the spacing
of the basis functions, the approximation error will likely be much
smaller than the error obtained from using a model that does not fit
the data perfectly and from estimating the parameters from the data.
Thus, for practical applications, the error in covariance induced by
the Hilbert space approximation technique will likely not matter much.
A more important consequence for practical applications when the
piecewise linear basis is used is that the Kriging estimation of the
field between two nodes in the triangulation is a linear interpolation
of the values at the nodes. Thus, variations on a scale smaller than
the spacing between the basis functions will not be captured correctly
in the Kriging prediction. For practical applications, it is therefore
often best to choose the number of basis functions depending on the
scale one is interested in the Kriging prediction on.

For the ozone data in Section \ref{paperC:sec:ozone}, the goal is to
estimate daily maps of global ozone. As we are not interested in
modeling small scale variations, we choose the number of basis
functions so that the mean distance between basis functions is about
258 km. For this basis, the smallest distance between two basis
functions is 222 km, and the largest distance is about 342 km.

\begin{figure}[t]

\includegraphics{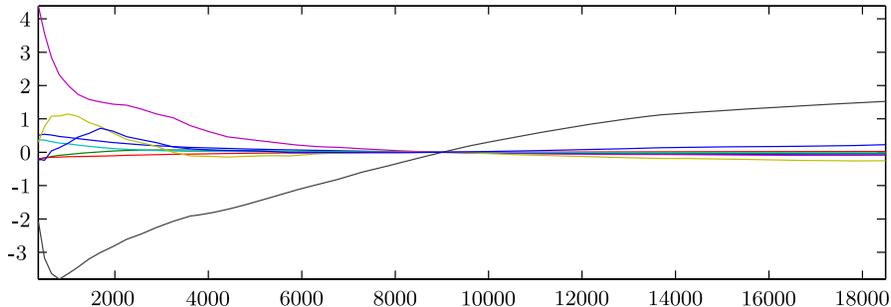}

\caption{Parameter estimates for the covariance parameters in model
F$'$
for the ozone data as functions of the number of basis functions in the
Hilbert space approximations.}
\label{fig:param_ests}
\end{figure}

Estimating the model parameters using different numbers of basis
functions will give different estimates, as the parameters are
estimated to maximize the likelihood for the approximate model instead
of the exact SPDE. An example of this can be seen in Figure \ref
{fig:param_ests} where the estimates of the covariance parameters for
model F' (see Section \ref{paperC:sec:ozone} for a model description)
for the ozone data are shown for varying numbers of basis functions.
Instead of showing the actual parameter estimates, the figure shows the
differences between the estimates and the estimate when using the basis
shown in Figure \ref{paperC:fig:triangulation}, which has 9002 basis
functions. Increasing the number of basis functions further, the
estimates will finally converge to the estimates one would get using
the exact SPDE representation. The curve that has not converged
corresponds to the dominating parameter in the vector field. Together
with $\kappa$, this parameter controls the correlation range of the
ozone field.

\section{Parameter estimation}\label{paperC:sec:paramest}
In this section a parameter estimation procedure for the
nested SPDE models is presented. One alternative would be to use a
Metropolis--Hastings algorithm, which is easy to implement, but
computationally inefficient. A better alternative is to use direct
numerical optimization to estimate the parameters.

Let $Y(\mathbf{s})$ be an observation of the latent field, $X(\mathbf{s})$,
given by \eqref{paperC:eq:model3} or \eqref{paperC:eq:modelnonstat},
under mean zero Gaussian measurement noise, $\mathcal{E}(\mathbf
{s})$, with
variance~$\sigma^2$:
%
\begin{equation}
\label{paperC:eq:measurement}
Y(\mathbf{s}) = X(\mathbf{s}) + \mathcal{E}(\mathbf{s}).
\end{equation}
Using the approximation procedure from Section
\ref{paperC:sec:galerkin}, and assuming a regression model for the
latent field's mean value function, $\mu(\mathbf{s})$, the measurement
equation can then be written as
\begin{eqnarray*}
\mathbf{Y} = \mathbf{M}\bolds{\mu} + \bolds{\Phi}\mathbf{w} +
\bolds{\varepsilon},
\end{eqnarray*}
where $\mathbf{M}$ is a matrix with the regression basis functions
evaluated at
the measurement locations, and $\bolds{\mu}$ is a vector containing the
regression
coefficients that have to be estimated. The matrix $\bolds{\Phi}$
contains the basis functions for the Hilbert space approximation
procedure evaluated at the measurement locations, and $\mathbf{w}$ is the
vector with the
stochastic weights. In Section~\ref{paperC:sec:galerkin} it was shown
that the vector $\mathbf{w}$ is Gaussian with mean zero and covariance matrix
$\mathbf{H}\mathbf{Q}_{X_0}^{-1}\mathbf{H}^{\top}$. Both $\mathbf
{Q}_{X_0}$ and
$\mathbf{H}$
are sparse matrices, but neither the covariance matrix nor the
precision matrix for $\mathbf{w}$ is sparse. Thus, it would seem as
if one had to work with a dense covariance matrix, which would make
maximum likelihood parameter estimation computationally infeasible for
large data sets. However, because of the product form of the
covariance matrix, one has that $\mathbf{w} = \mathbf{H}\mathbf
{w}_0$, where
$\mathbf{w}_0 \sim\mathsf{N}_{C}(\mathbf{0},\mathbf{Q}_{X_0})$.
Hence, the
observation equation
can be rewritten as
%
\begin{equation}\label{paperC:eq:obs}
\mathbf{Y} = \mathbf{M}\bolds{\mu} + \bolds{\Phi} \mathbf{H}
\mathbf{w}_0 + \bolds{\varepsilon}.
\end{equation}
Interpreting $\bolds{\Lambda} = \bolds{\Phi} \mathbf{H}$ as an
observation
matrix that depends on some of the parameters in the model, $\mathbf
{Y} -
\mathbf{M}\bolds{\mu}$ can now be seen as noisy observations of
$\mathbf{w}_0$,
which has a sparse
precision matrix. The advantage with using \eqref{paperC:eq:obs} is
that one then is in the setting of having observations of a latent
Gaussian Markov random field, which facilitates the usage of sparse
matrix techniques in the parameter estimation.

Let $\bolds{\psi}$ denote all parameters in the model except for
$\bolds{\mu}$. Assuming that $\bolds{\mu}$ and $\bolds{\psi}$
are a priori
independent, the posterior density can be written as
\begin{eqnarray*}
\pi(\mathbf{w}_0,\bolds{\mu},\bolds{\psi} | \mathbf{Y})
\propto\pi(\mathbf{Y}|\mathbf{w}_0,\sigma^2)\pi(\mathbf
{w}_0|\bolds{\mu},\bolds{\psi})\pi(\bolds{\mu})\pi(\bolds
{\psi}).
\end{eqnarray*}
Using a Gaussian prior distribution with mean $\bolds{\mu}$ and precision
$\mathbf{Q}_{\mu}$ for the mean parameters, the posterior distribution
can be reformulated as
%
\begin{equation}\label{paperC:eq:post}
\pi(\mathbf{w}_0,\bolds{\mu},\bolds{\psi} | \mathbf{Y})
\propto\pi(\mathbf{w}_0|\bolds{\mu}, \bolds{\psi}, \mathbf
{Y})\pi(\bolds{\mu} |
\bolds{\psi}, \mathbf{Y})\pi(\bolds{\psi}|\mathbf{Y}),
\end{equation}
where $\mathbf{w}_0|\bolds{\mu}, \bolds{\psi}, \mathbf{Y} \sim
\mathsf{N}_{C}(\mathbf{b},\hat{\mathbf{Q}})$, $\bolds{\mu} |
\bolds{\psi}, \mathbf{Y}
\sim
\mathsf{N}_{C}(\mathbf{b}_{\mu},\hat{\mathbf{Q}}_{\mu})$, and
\begin{eqnarray*}
\mathbf{b} &=& \frac{1}{\sigma^2}\bolds{\Lambda}^{\top}(\mathbf
{Y}-\mathbf{M}\bolds{\mu}),\qquad
\mathbf{b}_{\mu} = \mathbf{Q}_{\mu}\mathbf{m}_{\mu} +
\frac{\mathbf{M}^{\top}\mathbf{Y}}{\sigma^2} - \frac{\mathbf
{M}^{\top}
\bolds{\Lambda}\hat{\mathbf{Q}}^{-1} \bolds{\Lambda}^{\top
}\mathbf{Y}}{\sigma
^4},
\\
\hat{\mathbf{Q}} &=& \mathbf{Q}_{w_0} + \frac{1}{\sigma^2}\bolds
{\Lambda}^{\top}
\bolds{\Lambda},\qquad
\hat{\mathbf{Q}}_{\mu} = \mathbf{Q}_{\mu} + \frac{\mathbf
{M}^{\top}\mathbf{M}}{\sigma
^2} -
\frac{\mathbf{M}^{\top} \bolds{\Lambda}\hat{\mathbf{Q}}^{-1}
\bolds{\Lambda}^{\top
}\mathbf{M}}{\sigma^4}.
\end{eqnarray*}
The calculations are omitted here since these expressions are
calculated similarly to the posterior reformulation in
\citet{lindstrom092}, which gives more computational details. Finally,
the marginal posterior density $\pi(\bolds{\psi} | \mathbf{Y})$
can be shown
to be
\begin{eqnarray*}
\pi(\bolds{\psi} | \mathbf{Y}) \propto
\frac{|\mathbf{Q}_{w_0}|^{{1/2}}\pi(\bolds{\psi})}{|\hat
{\mathbf{Q}}|^{
{1/2}} |\hat{\mathbf{Q}}_{\mu}|^{{1/2}} |\sigma
\mathbf{I}|}\exp\biggl(\frac{1}{2\sigma^2}\mathbf{Y}^{\top}\biggl(\frac{
\bolds{\Lambda}\hat
{\mathbf{Q}}^{-1} \bolds{\Lambda}^{\top}}{\sigma^2} - \mathbf
{I}\biggr)\mathbf{Y} +
\frac{\mathbf{b}_{\mu}^{\top}\hat{\mathbf{Q}}_{\mu}^{-1}\mathbf
{b}_{\mu
}}{2} \biggr).
\end{eqnarray*}
By rewriting the posterior as \eqref{paperC:eq:post}, it can be
integrated with respect to~$\mathbf{w}_0$~and $\bolds{\mu}$, and
instead of
optimizing the full posterior with respect to $\mathbf{w}_0$, $\bolds
{\mu}$
and~$\bolds{\psi}$,~on\-ly the marginal posterior
$\pi(\bolds{\psi}|\mathbf{Y})$ has to be optimized with respect to
$\bolds{\psi}$. This is a lower dimensional optimization problem, which
substantially decreases the computational complexity. Given the
optimum, $\bolds{\psi}_{\mathrm{opt}} = \operatorname
{argmax}_{\bolds{\psi}}\pi
(\bolds{\psi}|\mathbf{Y})$,
$\bolds{\mu}_{\mathrm{opt}}$ is then given by $\bolds{\mu}_{\mathrm
{opt}} =
\hat{\mathbf{Q}}_{\mu}^{-1}\mathbf{b}_{\mu}$. In practice, the numerical
optimization is carried out on $\log\pi(\bolds{\psi} | \mathbf{Y})$.

\subsection{Estimating the parameter uncertainty}
There are several ways one could estimate the uncertainty in the
parameter estimates obtained by the parameter estimation procedure
above. The simplest estimate of the uncertainty is obtained by
numerically estimating the Hessian of the marginal posterior evaluated
at the estimated parameters. The diagonal elements of the inverse of
the Hessian can then be seen as estimates of the variance for the
parameter estimates.

Another method for obtaining more reliable uncertainty estimates is to
use a Metropolis--Hastings based MCMC algorithm with proposal kernel
similar to the one used in \citet{lindstrom092}. A quite efficient
algorithm is obtained by using random walk proposals for the
parameters, where the correlation matrix for the proposal distribution
is taken as a rescaled version of the inverse of the Hessian matrix
[\citet{gelman96}].

Finally, a third method for estimating the uncertainties is to use the
INLA framework [\citet{rue09}], available as an R package
(\url{http://www.r-inla.org/}). In settings with latent Gaussian Markov
random fields, integrated nested Laplace approximations (INLA) provide
close approximations to posterior densities for a fraction of the cost
of MCMC. For models with Gaussian data, the calculated densities are
for practical purposes exact. In the current implementation of the INLA
package, handling the full nested SPDE structure is cumbersome, so
further enhancements are needed before one can take full advantage of
the INLA method for these models.

\subsection{Computational complexity}
In this section some details on the computational complexity for the
parameter estimation and Kriging estimation are given.

The most widely used method for spatial prediction is linear
Kriging. In the Bayesian setting, the Kriging predictor simply is the
posterior expectation of the latent field $X$ given data and the
estimated parameters. This expectation can be written as
\begin{eqnarray*}
\mathsf{E}(X|\bolds{\psi},\bolds{\mu}, \mathbf{Y}) = \mathbf
{M}\bolds{\mu} + \bolds{\Phi} \mathbf{H}
\mathsf{E}(\mathbf{w}_0) = \mathbf{M}\bolds{\mu} + \bolds{\Phi
} \mathbf{H} \hat{\mathbf{Q}}^{-1}\mathbf{b}.
\end{eqnarray*}
The computationally demanding part of this expression is to calculate
$\hat{\mathbf{Q}}^{-1}\mathbf{b}$. Since the $n \times n$ matrix
$\mathbf{Q}$ is
positive definite, this is most efficiently done using Cholesky
factorization, forward substitution and back substitution:
Calculate the Cholesky triangle $\mathbf{L}$ such that $\hat{\mathbf
{Q}} =
\mathbf{L}\mathbf{L}^{\top}$, and given $\mathbf{L}$, solve the
linear system
$\mathbf{L}\mathbf{x} = \mathbf{b}$. Finally, given $\mathbf{x}$, solve
$\mathbf{L}^{\top}\mathbf{y} = \mathbf{x}$, where now $\mathbf{y}$
satisfies $\mathbf{y} = \hat{\mathbf{Q}}^{-1}\mathbf{b}$.
Solving the forward substitution and back substitution are much less
computationally demanding than calculating the Cholesky
triangle. Hence, the computational cost for calculating the Kriging
prediction is determined by the cost for calculating $\mathbf{L}$.

The computational complexity for the parameter estimation is
determined by the optimization method that is used and the
computational complexity for evaluating the marginal log-posterior
$\log\pi(\bolds{\psi} | \mathbf{Y})$. The most computationally demanding
terms in
$\log\pi(\bolds{\psi} | \mathbf{Y})$ are the two log-determinants
$\log
|\mathbf{Q}_{w_0}|$
and $\log|\hat{\mathbf{Q}}|$ and the quadratic form
$\mathbf{Y}^{\top}\bolds{\Lambda}\hat{\mathbf{Q}}^{-1}
\bolds{\Lambda}\mathbf{Y}$, which are also most efficiently
calculated using Cholesky factorization. Given the Cholesky triangle
$\mathbf{L}$, the quadratic form can be obtained as
$\mathbf{x}^{\top}\mathbf{x}$, where $\mathbf{x}$ is the solution to
$\mathbf{L}\mathbf{x} = \bolds{\Lambda}\mathbf{Y}$, and the
log-determinant
$\log|\hat{\mathbf{Q}}|$ is simply the sum\footnote{Since only the
difference between the log-determinants is needed, one should implement
the calculation as $2\sum_{i=1}^n (\log L_{(i)}^{w_0} - \log\hat
{L}_{(i)})$, where $L_{(i)}^{w_0}$ and $\hat{L}_{(i)}$ are the diagonal
elements of the Cholesky factors, sorted in ascending order, and the
sum is ordered by increasing absolute values of the differences. This
reduces numerical issues.} $2\sum_{i=1}^n
\log\mathbf{L}_{ii}$. Thus, the computational cost for one evaluation
of the margi\-nal posterior is also determined by the cost for
calculating~$\mathbf{L}$. Because of the sparsity structure of $\hat
{\mathbf{Q}}$, this computational cost is $\mathcal{O}(n)$, $\mathcal{O}
(n^{3/2})$ and
$\mathcal{O}(n^2)$ for problems in one, two and three dimensions respectively
[see \citet{rue1} for more details].

The computational complexity for the parameter
estimation is highly dependent on the optimization method. If a
Broyden--Fletcher--Goldfarb-Shanno (BFGS) procedure is used without an
analytic expression for the gradients, the marginal posterior has to be
evaluated $p$ times for each step in the optimization, where $p$ is the
number of covariance parameters in the model. Thus, if $p$ is large and
the initial value for the optimization is chosen far from the optimal
value, many thousand
evaluations of the marginal posterior may be needed in the
optimization.

\section{Application: Ozone data}\label{paperC:sec:ozone}
On October 24, 1978, NASA launched the near-polar, Sun-synchronous
orbiting satellite Nimbus-7. The satellite carried a TOMS instrument
with the purpose of obtaining high-resolution global maps of
atmospheric ozone [\citet{toms1}]. The instrument measured backscattered
solar ultraviolet radiation at 35 sample points along a line
perpendicular to the orbital plane at 3-degree intervals from 51
degrees on the right side of spacecraft to 51 degrees on the left. A
new scan was started every eight seconds, and as the measurements
required sunlight, the measurements were made during the sunlit
portions of the orbit as the spacecraft moved from south to north. The
data measured by the satellite has been calibrated and
preprocessed into a ``Level 2'' data set of spatially and temporally
irregular Total Column Ozone (TCO) measurements following the satellite
orbit. There is also a daily ``Level 3'' data set with values processed
into a regular latitude-longitude grid. Both Level 2 and Level 3 data
have been
analyzed in recent papers in the statistical literature
[\citet{cressie08}, \citet{jun08}, \citet{stein07}].

In what follows, the nested SPDE models are used to obtain statistical
estimates of a daily ozone map using a part of the Level 2 data. In
particular, all data available for October 1st, 1988 is used, which is
the same data set that was used by \citet{cressie08}.

\subsection{Statistical model}
The measurement model \eqref{paperC:eq:measurement} is used for the
ozone data. That is, the measurements, $Y(\mathbf{s})$, are assumed to be
observations of a latent field of TCO ozone,
$X(\mathbf{s})$, under Gaussian measurement noise $\mathcal
{E}(\mathbf{s})$
with a constant variance $\sigma^2$. We let $X(\mathbf{s})$ have some mean
value function, $\mu(\mathbf{s})$, and let the covariance structure be
determined by a nested SPDE model. Inspired by \citet{jun08}, who
proposed using
differentiated Mat\'{e}rn fields for modeling TCO ozone, we use the
simplest nested SPDE model. Thus, $Z(\mathbf{s}) =
X(\mathbf{s}) - \mu(\mathbf{s})$ is generated by the system
\begin{eqnarray*}
\bigl(\kappa^2(\mathbf{s})-\Delta\bigr) Z_0(\mathbf{s}) &=& \mathcal
{W}(\mathbf{s})
\\
Z(\mathbf{s}) &=& \bigl( b(\mathbf{s}) + \mathbf{B}(\mathbf{s})^{\top
}\nabla\bigr)Z_0(\mathbf{s}) ,
\end{eqnarray*}
where $\mathcal{W}(\mathbf{s})$ is Gaussian white noise on the
sphere. If
$\kappa(\mathbf{s})$ is assumed to be constant, the ozone is modeled
as a
Gaussian field with a covariance structure that is obtained by applying
the differential operator $ ( b(\mathbf{s}) + \mathbf{B}(\mathbf
{s})^{\top
}\nabla
)$ to a stationary Mat\'{e}rn field, which is similar to the model by
\citet{jun08}. If, on the other hand, $\kappa$ is spatially varying,
the range of the Mat\'{e}rn-like covariance function can vary with
location. As in \citet{stein07} and \citet{jun08}, the mean can be
modeled using a regression basis of spherical harmonics; however, since
the data set only contains measurements from one specific day,
it is not possible to identify which part of the variation in the data
that comes from a varying mean and which part that can be explained by the
variance--covariance structure of the latent field. To avoid this
identifiability problem, $\mu(\mathbf{s})$ is assumed to be unknown but
constant. The parameter $\kappa^2(\mathbf{s})$ has to be positive,
and for
identifiability reasons, we also require $b(\mathbf{s})$ to be
positive. We, therefore, let $\log\kappa^2(\mathbf{s}) =
\sum_{k,m}\kappa_{k,m}Y_{k,m}(\mathbf{s})$ and $\log b(\mathbf{s}) =
\sum_{k,m}b_{k,m}Y_{k,m}(\mathbf{s})$, where $Y_{k,m}$ is the spherical
harmonic of order $k$ and mode $m$. Finally, the vector field
$\mathbf{B}(\mathbf{s})$ is modeled using the vector spherical
harmonics basis
functions $\Upsilon_{k,m}^{1}$ and $\Upsilon_{k,m}^2$, presented in
\hyperref[paperC:sec:vectorbasis]{Appendix}:
\begin{eqnarray*}
\mathbf{B}(\mathbf{s}) = \sum_{k,m}\bigl(B_{k,m}^1 \Upsilon
_{k,m}^1(\mathbf{s})+B_{k,m}^2
\Upsilon_{k,m}^2(\mathbf{s})\bigr).
\end{eqnarray*}

\begin{table}[b]
\caption{Maximal orders of the spherical harmonics used in the bases
for the different parameters and total number of covariance
parameters in the different models for $X(\mathbf{s})$}\label{paperC:tab1}
\begin{tabular*}{\tablewidth}{@{\extracolsep{4in minus 4in}}lccccccccccccc@{}}
\hline
& \textbf{A} & \textbf{B} & \textbf{C} & \textbf{D} & \textbf{E} & \textbf{F} & \textbf{G} & \textbf{H} & \textbf{I} & \textbf{J} & \textbf{K} & \textbf{L} & \textbf{M}\\
\hline
$\kappa^2(\mathbf{s})$ & 0 & 1 & \phantom{0}0 & \phantom{0}1 & \phantom{0}2 & \phantom{0}0 & \phantom{0}3 & \phantom{0}2 & \phantom{0}0 & \phantom{0}4 & \phantom{0}3 &
\phantom{0}0 & \phantom{0}4\\
$b(\mathbf{s})$ & 0 & 1 & \phantom{0}1 & \phantom{0}1 & \phantom{0}2 & \phantom{0}2 & \phantom{0}3 & \phantom{0}2 & \phantom{0}3 & \phantom{0}4
& \phantom{0}3 & \phantom{0}4 & \phantom{0}4\\
$\mathbf{B}(\mathbf{s})$ & 0 & 0 & \phantom{0}1 & \phantom{0}1 & \phantom{0}0 & \phantom{0}2 & \phantom{0}0 & \phantom{0}2 & \phantom{0}3 & \phantom{0}0 & \phantom{0}3
& \phantom{0}4 & \phantom{0}4\\[3pt]
Total & 2 & 8 & 11 & 14 & 18 & 26 & 32 & 34 & 47 & 50 & 62 & 75
& 98\\
\hline
\end{tabular*}
\tabnotetext[]{}{\textit{Notes}: The actual
number of
basis functions for $\kappa^2(\mathbf{s})$ and $b(\mathbf{s})$ are
given by
$(\mathit{ord} + 1)^2$, and for $\mathbf{B}(\mathbf{s})$, the actual number is
$2(\mathit{ord}+1)^2-2$, where $\mathit{ord}$ is the maximal order indicated in the table.}
\end{table}

To choose the number of basis functions for the parameters
$\kappa^2(\mathbf{s})$, $b(\mathbf{s})$ and $\mathbf{B}(\mathbf
{s})$, some model selection
technique has to be used. Model selection for this model class is
difficult since the models can have both nonstationary mean value
functions and nonstationary covariance structures. This makes
standard variogram techniques inadequate in general, and we instead
base the model selection on Akaike's Information Criterion (AIC) and
the Bayesian
Information Criterion (BIC) [\citet{learning}], which are
suitable model
selection
tools for the nested SPDE models since the likelihood for the data can
be evaluated efficiently.

We estimate 13 models with different numbers of covariance parameters,
presented in Table \ref{paperC:tab1}. The simplest model is a
stationary Mat\'{e}rn model, with four parameters to estimate,
and the most complicated model has $100$ parameters to estimate,
including the mean and the measurement noise variance. There
are three different types of models in Table \ref{paperC:tab1}: In the
first type (models B, E, G and J), $\kappa^2$ and $b$ are spatially
varying and the vector field $\mathbf{B}$ is assumed to be zero. In the
second type (models C, F, I and L), $b$ and $\mathbf{B}$ are spatially
varying and $\kappa^2$ is assumed to be constant. Finally, in the
third type (model D, H, K and M), all parameters are spatially varying.

\begin{figure}[t]

\includegraphics{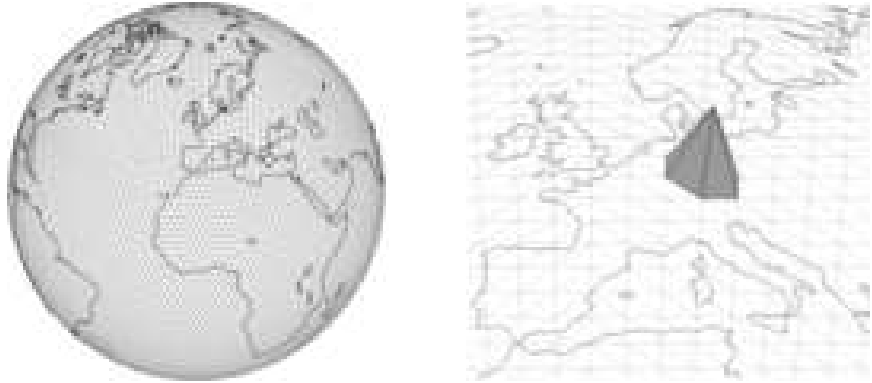}

\caption{The left part shows the triangulation of the Earth used to
define the piecewise linear basis functions in the Hilbert space
approximation for ozone data. Each basis function is one at a node in
the triangulation, and decreases linearly to zero at the neighboring
nodes. The right part of the figure shows one of these functions.}
\label{paperC:fig:triangulation}
\end{figure}

A basis of 9002 piecewise linear functions induced by a triangulation
of the Earth (see Figure~\ref{paperC:fig:triangulation}) is used in
the approximation procedure from \mbox{Section~\ref
{paperC:sec:galerkin}} to
get efficient representations of each model, and the parameters are
estimated using the procedure from Section~\ref{paperC:sec:paramest}.
The computational cost for the parameter estimation only depends on
the number of basis functions in the Hilbert space approximation, and
not on the number of data points, which makes inference efficient even
for this large data set.

\begin{figure}[t]

\includegraphics{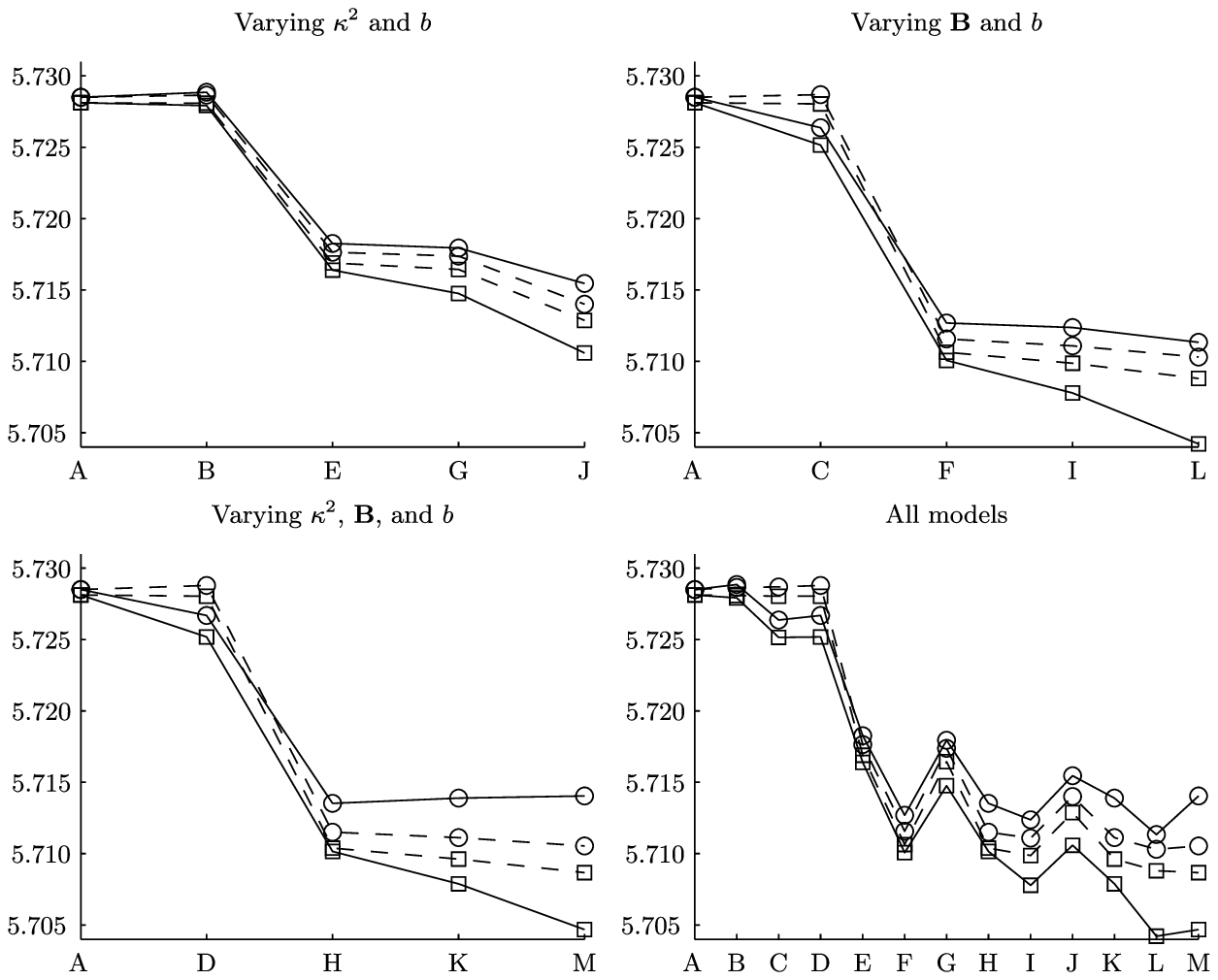}

\caption{AIC (squares) and BIC (circles) for the models A--M (solid
lines) and the axially symmetric models A$'$--M$'$ (dashed lines), scaled by
a factor $10^{-5}$. Note that the major improvement in AIC and BIC
occurs when the orders of the basis functions are increased from one to
two, and that the model type with spatially varying $b$ and $\mathbf{B}$
seems to be most appropriate for this data. Also note that the axially
symmetric model F$'$ is surprisingly good considering that it only has
$8$ covariance parameters.}
\label{paperC:fig:AIC}
\end{figure}

AIC and BIC for each of the fitted models can be
seen in Figure~\ref{paperC:fig:AIC}. The figure contains one panel for
each of
the three model types and one panel where AIC and BIC are shown for
all models at once. The major improvement in AIC and BIC occurs when
the orders of the basis functions are increased from one to
two. For the first model type, with spatially varying $\kappa^2$ and~$b$, the
figure indicates that the results could be improved by increasing the
orders of the basis functions further. However, for a given order of
the basis functions, the other two model types have much lower AIC and
BIC. Also, by comparing AIC and BIC for the second and third model
types, one finds that there is not much gain in letting $\kappa^2$ be
spatially varying. We therefore conclude that a model with spatially
varying $b$ and $\mathbf{B}$ is most appropriate for this data.

\begin{figure}[t]

\includegraphics{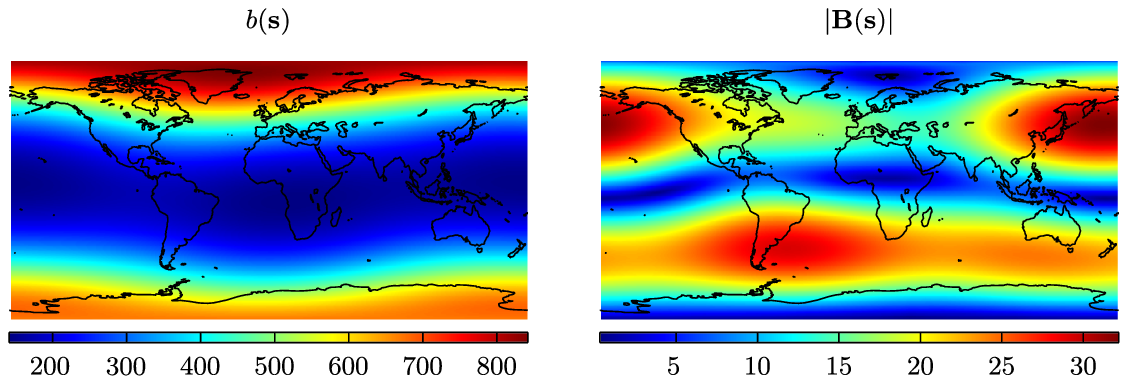}

\caption{Estimated variance-scaling parameter, $b(\mathbf{s})$, and
the the
norm of the vectors in the estimated vector field $\mathbf{B}(\mathbf
{s})$ for
model F. Note that the
estimates are fairly constant with respect to longitude, which
indicates that the latent field could be axially symmetric.}
\label{paperC:fig:parameters}
\end{figure}

The estimated parameters $b(\mathbf{s})$ and the length of the vectors
$\mathbf{B}(\mathbf{s})$ for model F are shown in Figure~\ref{paperC:fig:parameters}. One thing to note in this figure is
that the two parameters are fairly constant with respect to longitude, which
indicates that the latent field could be axially symmetric, an
assumption that was made by both \citet{stein07} and \citet{jun08}.
If the latent field indeed was axially symmetric, one would only need the
basis functions that are constant with respect to longitude in the
parameter bases. Since there is only one axially symmetric spherical
harmonic for each order, this assumption drastically reduces the
number of parameters for the models in Table \ref{paperC:tab1}. Let
A$'$--M$'$ denote the axially symmetric versions of models A--M. For these
models, the number of basis functions for both $\kappa^2(\mathbf{s})$ and
$b(\mathbf{s})$ is $\mathit{ord} + 1$, and the number of basis functions
for $\mathbf{B}(\mathbf{s})$ is $2(\mathit{ord}+1)-2$, where $\mathit{ord}$
is the
maximal order indicated in Table \ref{paperC:tab1}. The dashed lines
in Figure~\ref{paperC:fig:AIC} show AIC and BIC calculated for these
models. Among the axially symmetric models, model F$'$ is
surprisingly good considering that it only has $8$ covariance
parameters.

\begin{figure}[b]

\includegraphics{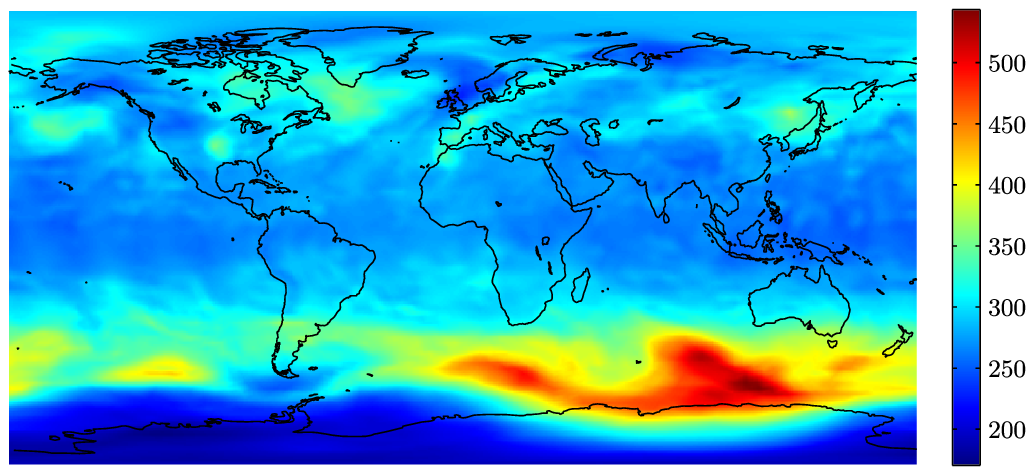}

\caption{Kriging estimate of TCO ozone in Dobson units using model F$'$.}
\label{paperC:fig:kriging}
\end{figure}

\begin{figure}[t]

\includegraphics{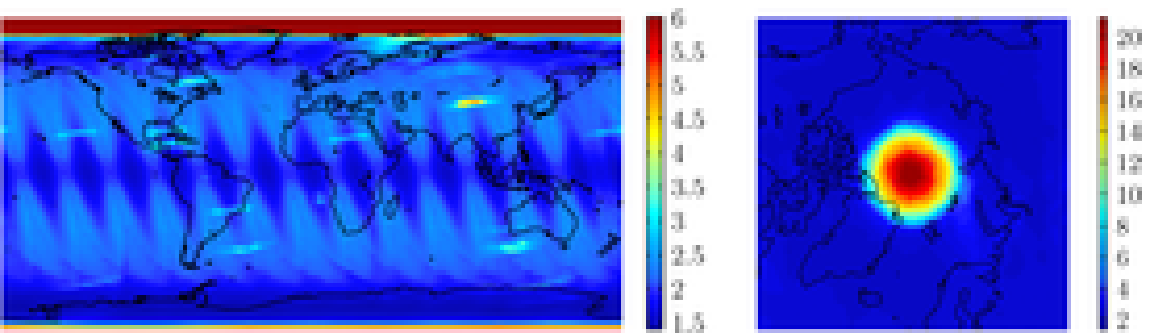}

\caption{Standard error in Dobson units for the Kriging estimate. The
color bar in the left part of the figure has been truncated at 6 Dobson
units. The behavior near the north pole can be seen in the right
part of the figure.}
\label{paperC:fig:krigingerror}
\end{figure}

\begin{figure}[b]

\includegraphics{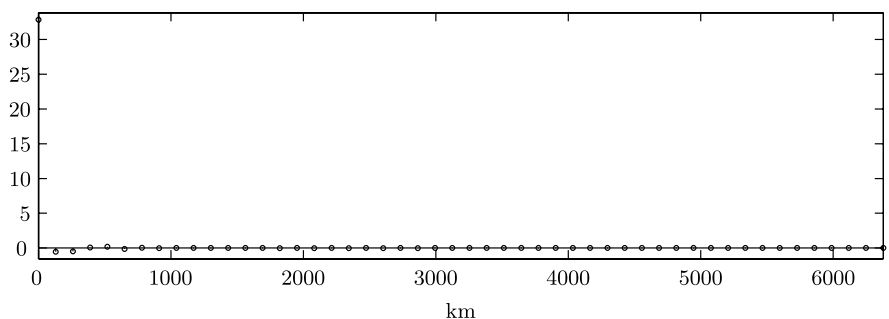}

\caption{Estimated covariance function for the Kriging residuals using
model F$'$.}
\label{fig:err_cov}
\end{figure}

\begin{figure}[t]

\includegraphics{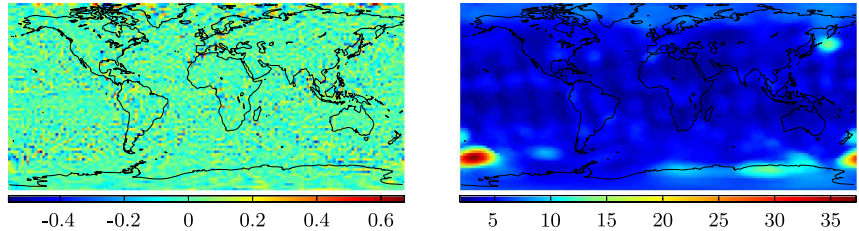}

\caption{Estimates of the local mean (left) and standard deviation
(right) for the Kriging residuals using model F$'$. The mean is fairly
constant across the globe, whereas the standard deviation is higher
close to the poles and at the international date line because of the
temporal structure in the data.}
\label{fig:err_var}
\end{figure}

The Kriging estimate and its standard error for model F$'$ are shown in
Figures \ref{paperC:fig:kriging} and  \ref{paperC:fig:krigingerror}
respectively. The oscillating behavior near
the equator for the standard error is explained by the fact that the
satellite tracks are furthest apart there, which results in sparser
measurements between the different tracks. Because the measurements are collected using backscattered sunlight, the
variance close to the north pole is high, as there are no measurements
there.
As seen in Figure \ref{fig:err_cov}, there is not much spatial
correlation in the residuals $\hat{\mathbf{X}} - \mathbf{Y}$, which
indicates a
good model fit. In~Figure \ref{fig:err_var}, estimates of the local
mean and variance of the residuals are shown. The mean is fairly
constant across the globe, but there is a slight tendency for higher
variance closer to the poles. This is due to the fact that the data
really is space--time data, as the measurements are collected during a
24-hour period. Since the different satellite tracks are closest near
the poles, the temporal variation of the data is most prominent here,
and especially near the international date line where data is collected
both at the first satellite track of the day and at the last track, 24
hours later. The area with high residual variance is one of those
places where measurements are taken both at the beginning and the end
of the time period, and where the ozone concentration has changed
during the time period between the measurements. One could include this
effect by allowing the variance of the measurement noise to be
spatially varying; however, one should really use a spatio-temporal
model for the data to correctly account for the effect, which is
outside the scope of this article.

\begin{table}[b]
\caption{Estimates of the covariance parameters in model F$'$ using all
data but the first track ($Y_f$), all data but the last track ($Y_l$),
and all data ($Y$)}
\label{paperC:tab2}
\begin{tabular*}{\tablewidth}{@{\extracolsep{4in minus 4in}}lccccccccc@{}}
\hline
& $\bolds{\kappa}$ & $\bolds{\sigma}$ & $\bolds{b_1}$ & $\bolds{b_2}$ & $\bolds{b_3}$ & $\bolds{B_1}$ & $\bolds{B_2}$ & $\bolds{B_3}$
& $\bolds{B_4}$\\
\hline
$Y_f$ & $0.74$ & $25.60$ & $5.85$ & $0.045$ & $0.34$ & $1.05$ & $2.59$
& $-6.84$ & $-0.84$\\

$Y_l$ & $0.73$ & $25.56$ & $5.82$ & $0.033$ & $0.34$ & $0.90$ & $2.38$
& $-7.01$ & $-0.82$\\

$Y$ & $0.67$ & $34.09$ & $5.75$ & $0.054$ & $0.36$ & $0.70$ & $2.48$ &
$-7.10$ & $-0.68$\\
\hline
\end{tabular*}
\end{table}

\begin{figure}[b]

\includegraphics{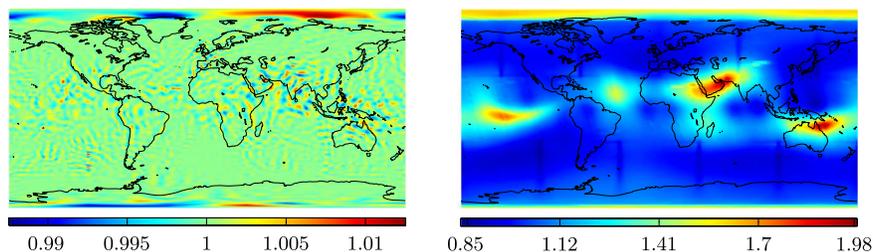}

\caption{The ratio between the kriging estimates using model F$'$ and
model M (left), and the ratio between the corresponding kriging
standard errors (right). Note that there is not much difference
between the Kriging estimates, whereas there is a clear difference
between the corresponding standard errors.}
\label{paperC:fig:krigdiff}
\end{figure}

To see how much the temporal structure near the international date line
influences the model fit, the parameters in model F$'$ are re-estimated
without using the first satellite track of the day and without using
the last track of the day. The estimated parameters can be seen in
Table \ref{paperC:tab2} and, as expected, the estimate of the
measurement noise variance is much lower when not using all date line
data. The estimates of the covariance parameters for the latent field
also change somewhat, but the large scale structure of the
nonstationarity is preserved.

To study how sensitive the Kriging estimates are to the model choice,
the ratio between the Kriging estimates for the simple model F$'$ and the
large model M, and the ratio between the corresponding Kriging standard
errors, are shown in Figure~\ref{paperC:fig:krigdiff}. There is not much difference between the
two Kriging estimates, whereas there is a clear difference between the
corresponding standard errors. Thus, if one only is
interested in the Kriging estimate, it does not matter much which model
is used, but if one also is interested in the standard error of
the estimate, the model choice greatly influences the results.

\subsection{Discussion}
Before the nested SPDE models were used on the ozone data, several
tests were performed on simulated data to verify that the model
parameters in fact could be estimated using the estimation procedure
in Section \ref{paperC:sec:paramest}.
These tests showed that the estimation procedure is robust given that
the initial values for the parameters are not chosen too far from
the true values. However, for nonstationary models with many
covariance parameters, it is not easy to choose the initial values.
To reduce this problem, the optimization is done in several steps. A
stationary Mat\'{e}rn model (model A) is estimated to get initial
values for
$\kappa_{0,0}$, $b_{0,0}$ and $\sigma^2$. To estimate model B, all
parameters are set to zero initially, except for the parameters that were
estimated in model A. Another layer of spherical harmonics is added to
the bases for $\kappa^2(\mathbf{s})$ and $b(\mathbf{s})$ for estimating
model E
using the model B
parameters as initial values. This step-wise procedure of adding layers
of spherical harmonics to the bases is then repeated to estimate the
larger models.
Numerical studies showed that this optimization procedure is quite
robust even for large models; however, as in most other numerical
optimization problems, there are no guarantees that the true optimal
values have been found for all models for the ozone data.

The application of the nested SPDE models to ozone data was inspired
by \citet{jun08}, who proposed using differentiated Mat\'{e}rn
fields for modeling TCO ozone, and we conclude this section with some
remarks on the similarities and differences between the nested SPDEs
and their models. The most general model in \citet{jun08} is on the form
\begin{eqnarray}\label{paperC:eq:jun}
X(\mathbf{s}) &=& P_1(l_2)X_0(\mathbf{s}) + \biggl(P_2(l_2)\frac{\partial
}{\partial l_2}
+P_3(l_2)\frac{\partial}{\partial l_1}\biggr)X_1(\mathbf{s})\nonumber
\\[-8pt]\\[-8pt]
&&{}+
P_4(l_2)\frac{\partial
}{\partial
l_1} X_2(\mathbf{s}),\nonumber
\end{eqnarray}
where $X_i, i=0,1,2$, are i.i.d. Mat\'{e}rn fields in $\mathbb{R}^3$, $P_i,
i=1,2,3,4$, are nonrandom functions depending on latitude, $l_1$
denoted longitude and $l_2$ denoted latitude. This model is similar to
the model used here, but there are some important differences. First
of all, \eqref{paperC:eq:jun} contains a sum of three independent
fields, which
we cannot represent since the approximation procedure in Section
\ref{paperC:sec:galerkin} in this case loses its computational
benefits. To get a model more similar to the nested SPDE model, one
would have
to let $P_4(l_2)\equiv0$, and $X_0(\mathbf{s}) = X_1(\mathbf{s})$.
Using $X_0
= X_1$
or $X_0$ and $X_1$ as i.i.d. copies of a Mat\'{e}rn field gives different
covariance functions, and without testing both cases it is
hard to determine what is more appropriate for ozone data.

Another important conceptual difference is how the methods deal with
the spherical topology. The Mat\'{e}rn fields in \citet{jun08} are
stochastic fields on $\mathbb{R}^3$, evaluated on the embedded sphere, which
is equivalent to using chordal distance as the metric in a regular
Mat\'{e}rn covariance function. One might instead attempt to evaluate
the covariance function using the arc-length distance, which is a more
natural metric on the sphere. However, Theorem~2 from \citet{gneiting98}
shows that for Mat\'{e}rn covariances with \mbox{$\nu\geq1$}, this
procedure does not generate positive definite covariance functions.
This means that the arc-length method cannot be used for any
differentiable Mat\'{e}rn fields. On the other hand, the nested SPDEs
are directly defined on the sphere, and therefore
inherently use the arc-length distance.

There is, in theory, no difference between writing the
directional derivative of $X(\mathbf{s})$ as $(P_2(l_2)\frac{\partial
}{\partial
l_2}
+P_3(l_2)\frac{\partial}{\partial l_1})X_1(\mathbf{s})$ or
$\mathbf{B}(\mathbf{s})^{\top}\nabla X(\mathbf{s})$, but the latter
is easier
to work
with in practice. If a vector field basis is used to model $\mathbf{B}(s)$,
the process will not have any singularities as long as the basis
functions are nonsingular, which is the case for the basis used in this
paper. If, on the other hand, $P_2(l_2)$ and $P_3(l_2)$ are modeled
separately, the process will be singular at the poles unless certain
restrictions on the two functions are met. This fact is indeed noted by
\citet{jun08}, but the authors do not seem to take the restrictions
into account in the parameter estimation, which causes all their
estimated models to have singularities at the poles.

Finally, the nested SPDE models are computationally efficient
also for spatially irregular data, which allowed us to work
with the TOMS Level 2 data instead of the gridded Level 3 data.

\section{Concluding remarks}\label{paperC:sec:conclusions}
There is a need for computationally efficient stochastic models for
environmental data. \citet{lindgren10} introduced an efficient
procedure for obtaining Markov approximations of, possibly
nonstationary, Mat\'{e}rn fields by considering Hilbert space
approximations of the SPDE
\begin{eqnarray*}
\bigl(\kappa(\mathbf{s})^2-\Delta\bigr)^{{\alpha/2}}X(\mathbf{s}) =
\phi(\mathbf{s})
\mathcal{W}(\mathbf{s}).
\end{eqnarray*}
In this work, the class of nonstationary nested SPDE models generated
by \eqref{paperC:eq:modelnonstat} was introduced, and it was shown how
the approximation methods in \citet{lindgren10} can be extended to this
larger class of models. This model class contains a wider family of
covariance models, including both Mat\'{e}rn-like covariance functions
and various oscillating covariance functions. Because of the additional
differential operator $\mathcal{L}_2$, the Hilbert space approximations
for the nested SPDE models do not have the Markov structure the model
in \citet{lindgren10} has, but all computational benefits from the
Markov properties are preserved for the nested SPDE models using the
procedure in Section \ref{paperC:sec:paramest}. This allows us to fit
complicated models with over 100 parameters to data sets with several
hundred thousand measurements using only a standard personal computer.

By choosing $\mathcal{L}_2 = b + \mathbf{B}^{\top}\nabla$, one
obtains a
model similar to what \citet{jun08} used to analyze TOMS Level 3 ozone
data, and we used this restricted nested SPDE model to analyze the
global spatially irregular TOMS Level 2 data. This application
illustrates the ability to use the model class to produce nonstationary
covariance models on general smooth manifolds which efficiently can be
used to study large spatially irregular data sets.

The most important next step in this work is to make a spatio-temporal
extension of the model class. This would allow us to produce not only
spatial but also spatio-temporal ozone models and increase the
applicability of
the model class to other environmental modeling problems where time
dependence is a necessary model component.

\begin{appendix}
\section*{Appendix: Vector spherical harmonics}\label{paperC:sec:vectorbasis}
When using the nonstationary model \eqref{paperC:eq:modelnonstat} in
practice, we assume that the parameters in the model can be expressed
in terms of some basis functions. If working on the sphere, spherical
harmonics is a convenient basis for the parameters taking values in
$\mathbb{R}$. On real form, the spherical harmonic $Y_{k,m}(\mathbf
{s})$ of order
$k\in\mathbb{N}_0$ and mode $m=-k, \ldots, k$ is defined as
\begin{eqnarray*}
\ Y_{k,m}(\mathbf{s}) =
\sqrt{\frac{2k+1}{4\pi}\cdot\frac{(k-|m|)!}{(k+|m|)!}}
\cdot
\cases{
\sqrt{2} \sin(ml_1) P_{k,-m}(\sin l_2), & \quad$-k\leq m <0,$\cr
P_{k,0}(\sin l_2), & \quad$m=0,$\cr
\sqrt{2} \cos(ml_1) P_{k,m}(\sin l_2), & \quad$0 < m \leq k,$
}
\end{eqnarray*}
where $l_2$ is the latitude, $l_1$ is the longitude, and
$P_{k,m}(\cdot)$ are associated Legendre functions. We,
however, also need a basis for the vector fields $\mathbf
{B}_{i}(\mathbf{s})$,
determining the direction and magnitude of differentiation. Since the
vector fields in each point on the sphere must lie in the tangent space
of $\mathbb{S}^2$, the basis functions also must satisfy this. A basis
with this property is obtained by using a subset of the vector
spherical harmonics [\citet{hill54}]. For each spherical harmonic
$Y_{k,m}(\mathbf{s})$, $k>0$,
define the two vector spherical harmonics
\begin{eqnarray*}
\Upsilon_{k,m}^1(\mathbf{s}) &=& \nabla_{\mathbb{S}^2}
Y_{k,m}(\mathbf{s}),
\\
\Upsilon_{k,m}^2(\mathbf{s}) &=& \nabla_{\mathbb{S}^2}
Y_{k,m(\mathbf{s})}
\times\mathbf{s}.
\end{eqnarray*}
Here $\times$ denotes the cross product in $\mathbb{R}^3$ and $\nabla
_{\mathbb{S}^2}$
is the gradient on $\mathbb{S}^2$. By\vspace*{1pt} defining the basis in this way, all
basis functions in $\Upsilon^1 = \{\Upsilon_{k,m}^1\}$ and $\Upsilon^2
= \{\Upsilon_{k,m}^2\}$ will obviously lie in the tangent space of
$\mathbb{S}^2$. It is also easy to see that the basis is orthogonal in the
sense that for any $k,l>0$, $-k\leq m \leq k$, and $-l\leq n \leq l$,
one has
\begin{eqnarray*}
\langle{\Upsilon_{k,m}^1},{\Upsilon_{l,n}^2} \rangle_{\mathbb
{S}^2} &=&
0,
\\
\langle{\Upsilon_{k,m}^1},{\Upsilon_{l,n}^1} \rangle_{\mathbb
{S}^2} &=&
k(k+1)\delta
_{k-l}\delta_{m-n},
\\
\langle{\Upsilon_{k,m}^2},{\Upsilon_{l,n}^2} \rangle_{\mathbb
{S}^2} &=&
k(k+1)\delta
_{k-l}\delta_{m-n}.
\end{eqnarray*}
These are indeed desirable properties for a vector field basis, but
for the basis to be of any use in practice, a method for calculating
the basis functions explicitly is needed. Such explicit expressions
are given in the following proposition.

\begin{prop}\label{prop:vsh}
With $\mathbf{s} = (x,y,z)^{\top}$, $\Upsilon_{k,m}^1(\mathbf{s})$ and
$\Upsilon_{k,m}^2(\mathbf{s})$ can be written as
\begin{eqnarray*}
\Upsilon_{k,m}^1(\mathbf{s}) &=& \frac{1}{1-z^2}\left[
\matrix{
-myY_{k,-m}(\mathbf{s}) - c_{k,m}xzY_{k-1,m}(\mathbf
{s})+kxz^2Y_{k,m}(\mathbf{s})
\cr
mxY_{k,-m}(\mathbf{s}) - c_{k,m}yzY_{k-1,m}(\mathbf
{s})+kyz^2Y_{k,m}(\mathbf{s})
\cr
c_{k,m}(1-z^2)Y_{k-1,m}(\mathbf{s}) - (1-z^2)kzY_{k,m}(\mathbf{s})
}
\right],
\\
\Upsilon_{k,m}^2(\mathbf{s}) &=& \frac{1}{1-z^2}\left[
\matrix{
kzyY_{k,m}(\mathbf{s})-c_{k,m}yY_{k-1,m}(\mathbf
{s})+mzxY_{k,-m}(\mathbf{s})\cr
-kxzY_{k,m}(\mathbf{s})+c_{k,m}xY_{k-1,m}(\mathbf
{s})+myzY_{k,-m}(\mathbf{s})\cr
-m(1-z^2)Y_{k,-m}(\mathbf{s})
}
\right],
\end{eqnarray*}
where
\begin{eqnarray*}
c_{k,m} = \sqrt{\frac{(2k+1)(k^2-|m|^2)}{2k-1}}.
\end{eqnarray*}
\end{prop}

\begin{pf}
One has that $ \nabla_{\mathbb{S}^2} Y_{k,m} = P_{\mathbb
{S}^2}(\nabla_{\mathbb{R}^3}
Y_{k,m})$, that is,~the gradient on $\mathbb{S}^2$ can be obtained by first
calculating the gradient in $\mathbb{R}^3$ and then projecting the
result onto
$\mathbb{S}^2$. If $c_k^m$ denotes the normalization constant for
the spherical harmonic $Y_{k,m}(\mathbf{s})$, and the recursive relation
\begin{eqnarray*}
(1-z^2)\frac{\partial}{\partial z}P_{k,m}(z) = kzP_{k,m}(z)-(k+m)P_{k-1,m}(z)
\end{eqnarray*}
is used, one has that
\begin{eqnarray*}
\frac{\partial}{\partial z}Y_{k,m}(\mathbf{s}) = \frac
{1}{1-z^2}\biggl(kzY_{k,m}(\mathbf{s})-(k+|m|)\frac
{c_k^m}{c_{k-1}^m}Y_{k-1,m}(\mathbf{s})\biggr).
\end{eqnarray*}
Now, using that $\tan(l_1) = x^{-1}y$, one has
\begin{eqnarray*}
\frac{\partial l_1}{\partial x} &=& -\cos^2(l_1)\frac{y}{x^2} =
-\frac{y}{1-z^2},
\\
\frac{\partial l_1}{\partial y} &=& \cos^2(l_1)\frac{1}{x} =
\frac{x}{1-z^2},
\end{eqnarray*}
where the last equalities hold on $\mathbb{S}^2$. Using these
relations gives
\begin{eqnarray*}
\frac{\partial}{\partial x}Y_{k,m}(\mathbf{s}) =
-\frac{my}{1-z^2}Y_{k,-m}(\mathbf{s}),\qquad
\frac{\partial}{\partial y}Y_{k,m}(\mathbf{s}) = \frac
{mx}{1-z^2}Y_{k,-m}(\mathbf{s}).
\end{eqnarray*}
Thus, with
\begin{eqnarray*}
c_{k,m} \mathrel{\triangleq}(k+|m|)\frac{c_k^m}{c_{k-1}^m} = \sqrt
{\frac
{(2k+1)(k^2-|m|^2)}{2k-1}},
\end{eqnarray*}
one has that
\begin{eqnarray*}
\nabla_{\mathbb{R}^3}Y_{k,m}(\mathbf{s}) = \frac{1}{1-z^2}\left[
\matrix{
- myY_{k,-m}(\mathbf{s})\cr
mxY_{k,-m}(\mathbf{s})\cr
kzY_{k,m}(\mathbf{s})-c_{k,m}Y_{k-1,m}(\mathbf{s})
}
\right].
\end{eqnarray*}
Finally, the desired result is obtained by calculating
\begin{eqnarray*}
\Upsilon_{k,m}^1 &=& \nabla_{\mathbb{S}^2}Y_{k,m} =
\mathbf{P}_{\mathbb{S}^2}\nabla_{\mathbb{R}^3}Y_{k,m},
\\
\Upsilon_{k,m}^2 &=& \Upsilon_{k,m}^1\times\mathbf{s} = \mathbf
{S}_{\times
}\Upsilon_{k,m}^1,
\end{eqnarray*}
where
\begin{eqnarray*}
\mathbf{P}_{\mathbb{S}^2} = (I-\mathbf{s}\mathbf{s}^{\top}) = \left[
\matrix{
1-x^2 & -xy & -xz\cr
-xy & 1-y^2 & -yz\cr
-xz & -yz & 1-z^2
}
\right],\qquad
\mathbf{S}_{\times} = \left[
\matrix{
0 & -z & y\cr
z & 0 & -x\cr
-y & x & 0
}
\right].\\[3pt]
\end{eqnarray*}
\upqed\end{pf}
\end{appendix}

%


\printaddresses

\end{document}